\documentclass[fleqn,usenatbib]{mnras}

\DeclareRobustCommand{\VAN}[3]{#2}
\let\VANthebibliography\thebibliography
\def\thebibliography{\DeclareRobustCommand{\VAN}[3]{##3}\VANthebibliography}

\usepackage{graphicx}	
\usepackage{amsmath}	
\usepackage{amssymb}	
\usepackage{pdflscape}	
\usepackage{caption}

\usepackage{pgfplotstable}
\usepackage{csvsimple}
\usepackage{booktabs} 

\usepackage{soul}

\newcommand{\Msun}{{$M_{\odot}$}}
\newcommand{\kms}{km\,s$^{-1}$}

\newcommand\arcdeg{\mbox{$^\circ$}}

\newcommand{\OIV}{[\ion{O}{iv}]}
\newcommand{\OIII}{[\ion{O}{iii}]}

\newcommand{\NII}{[\ion{N}{ii}]}

\newcommand{\SIV}{[\ion{S}{iv}]}
\newcommand{\SIII}{[\ion{S}{iii}]}

\usepackage{amssymb}

\makeatletter
\newcommand{\undersim}[1]{\mathrel{\mathpalette\@undersim{#1}}}
\newcommand{\@undersim}[2]{%
  \vcenter{%
    \ialign{%
      ##\cr
      $\m@th#1#2$\cr
      \noalign{\nointerlineskip\kern.2ex}
      $\m@th#1\sim$\cr
      \noalign{\kern-.4ex}
    }%
  }%
}
\newcommand{\gsim}{\undersim{>}}

\makeatother

\pgfplotsset{compat=1.18}

\title[]{\textit{JWST} observations of the Ring Nebula (NGC\,6720): I. Imaging of the rings, globules, and arcs
}
\author[Wesson et al.]{%
R. Wesson$^{1,2}$,
Mikako Matsuura$^1$,
Albert A. Zijlstra$^3$, 
Kevin Volk$^4$,
Patrick J. Kavanagh$^{5,6}$, \newauthor
Guillermo Garc\'{\i}a-Segura$^7$,   
I. McDonald$^{3,8}$, 
Raghvendra Sahai$^9$,
M. J. Barlow$^2$,
Nick L. J. Cox$^{10}$,  \newauthor 
Jeronimo Bernard-Salas$^{10,11}$, 
Isabel Aleman$^{12,13}$,
Jan Cami$^{14,15,16}$,
Nicholas Clark$^{14,15}$,\newauthor 
Harriet L. Dinerstein$^{17}$, 
K. Justtanont$^{18}$, 
Kyle F. Kaplan$^{17}$, 
A. Manchado$^{19,20,21}$,
Els Peeters$^{14,15,16}$, \newauthor
Griet C. Van de Steene$^{22}$,  
Peter A. M. van Hoof$^{22}$
\\
$^1$
Cardiff Hub for Astrophysics Research and Technology (CHART),
School of Physics and Astronomy, Cardiff University, \\
The Parade, Cardiff CF24 3AA, UK\\
$^2$ 
Department of Physics and Astronomy, University College London, Gower Street, London WC1E 6BT, United Kingdom\\
$^3$ 
Jodrell Bank Centre for Astrophysics, Department of Physics \&\ Astronomy, The University of Manchester, Oxford Road, \\
Manchester M13 9PL, UK \\
$^4$ 
Space Telescope Science Institute, 3700 San Martin Drive, Baltimore, MD 21218, USA\\
$^{5}$
Department of Experimental Physics, Maynooth University, Maynooth, Co Kildare, Ireland \\
$^{6}$
School of Cosmic Physics, Dublin Institute for Advanced Studies, 31 Fitzwilliam Place, Dublin 2, Ireland \\
$^7$
Instituto de Astronom\'ia, Universidad Nacional Aut\'onoma de M\'exico, Ensenada, Mexico\\
$^8$
Department of Physical Sciences, The Open University, Walton Hall, Milton Keynes, MK7 6AA, UK\\
$^9$ 
Jet Propulsion Laboratory, California Institute of Technology, Pasadena, CA, USA \\
$^{10}$ 
ACRI-ST, Centred’Etudes et de Recherche de Grasse (CERGA), 10Av. Nicolas Copernic, 06130 Grasse, France \\
$^{11}$ 
INCLASS Common Laboratory., 10 Av. Nicolas Copernic, 06130 Grasse, France \\
$^{12}$ Department of Computer Science, Institute
 of Mathematics and Statistics, University of S\~{a}o Paulo, Rua do Mat\~{a}o 1226, Cidade \\Universit\'{a}ria, 05508-090, S\~{a}o Paulo, SP, Brazil\\
$^{13}$ Instituto de F\'isica e Qu\'imica, Universidade Federal de Itajub\'a, Av. BPS 1303, Pinheirinho, 37500-903, Itajub\'a, MG, Brazil\\
$^{14}$
Department of Physics and Astronomy, University of Western Ontario, London, Ontario, Canada \\ 
$^{15}$
Institute for Earth and Space Exploration, University of Western Ontario, London, Ontario, Canada \\
$^{16}$
SETI Institute, Mountain View, CA, USA \\
$^{17}$
University of Texas at Austin, Austin, TX 78712, USA \\
$^{18}$
Chalmers University of Technology, Onsala Space Observatory, S-439 92 Onsala, Sweden \\
$^{19}$
Instituto de Astrofísica de Canarias, E-38205 La Laguna, Tenerife, Spain \\
$^{20}$
Departamento de Astrofísica, Universidad de La Laguna, E-38206 La Laguna, Tenerife, Spain \\
$^{21}$
Consejo Superior de Investigaciones Científicas (CSIC), Spain \\
$^{22}$
Royal Observatory of Belgium, Ringlaan 3, B-1180 Brussels, Belgium \\
}
\date{Accepted XXX. Received YYY; in original form ZZZ}

\pubyear{2023}

\begin{document}
\label{firstpage}
\pagerange{\pageref{firstpage}--\pageref{lastpage}}
\maketitle

\begin{abstract}

We present \textit{JWST} images of the well-known planetary nebula NGC\,6720 (the Ring Nebula), covering wavelengths from 1.6$\mu$m to 25 $\mu$m. The bright shell is strongly fragmented with some 20\,000 dense globules, bright in H$_2$, with a characteristic diameter of 0.2~arcsec and density $n_{\rm H} \sim 10^5$--$10^6$\,cm$^{-3}$. The  shell contains a thin ring of polycyclic aromatic hydrocarbon (PAH) emission. H$_2$ is found throughout the shell and in the halo. H$_2$ in the halo may be located on the swept-up walls of a biconal polar flow. The central cavity is shown to be filled with high ionization gas and shows two linear structures. The central star is located 2~arcsec from the emission centroid of the cavity and shell. Linear features (`spikes') extend outward from the ring, pointing away from the central star. Hydrodynamical simulations are shown which reproduce the clumping and possibly the spikes. Around ten low-contrast, regularly spaced concentric arc-like features are present; they suggest orbital modulation by a low-mass companion with a period of about 280~yr A previously known much wider companion is located at a projected separation of about 15\,000~au; we show that it is an  M2--M4 dwarf. The system is therefore a triple star. These features, including the multiplicity, are similar to those seen in the Southern Ring Nebula (NGC\,3132) and may be a common aspect of such nebulae. 
\end{abstract}

\begin{keywords}
    planetary nebulae: general --
    planetary nebulae: individual: NGC6720 --
    circumstellar matter
\end{keywords}



\section{Introduction}

Planetary nebulae (PNe) are composed of the ionized ejecta from low- and intermediate-mass stars ($<8$~\Msun) at the ends of their lives.  PNe can be used 
as astrophysical laboratories, having 
a single exciting central star, and hosting a range of ionized, neutral, and molecular lines as well as dust emission. They are ideal objects to study the physics and chemistry of 
gaseous media under well-defined conditions. The structures shown by PNe range from their large scale shape to small scale condensations, including filaments and globules. They can be used to study the hydrodynamical origin of such features. Specific questions that can be studied using PNe include the formation and destruction of molecules including H$_2$ and polycyclic aromatic hydrocarbons (PAHs), and the role of stellar interactions in the shaping of the nebulae.

We report here on \textit{JWST} imaging of the Ring Nebula, NGC 6720, using 13 filters  from 1.6~$\mu$m to 25$\mu$m. The different filters trace a range of emission lines and dust features. \textit{JWST}'s high angular resolution enables us to trace the ionized, molecular, and dust components of the nebula. Due to its proximity, high angular resolution and high sensitivity \textit{JWST} images can reveal details of the physics and chemistry in small structures such as globules and filaments in this PN. These features are shared with the Southern Ring Nebula (NGC 3132), which was also imaged by \textit{JWST} \citep{DeMarco2022}. 

This paper is one of four describing new \textit{JWST} observations of the Ring Nebula, together with a study of the central star and its close environs (Sahai et al. in prep), a study of the rich H$_2$ emission line spectra in two subregions of the nebula (van Hoof et al. in prep), and a study of the PAH emission in two subregions of the nebula (Clark et al., in prep.).


\begin{table*}
\begin{center}
\caption{Observing log \label{observing_log} }
\begin{filecontents*}{observing_log.csv}
Instr.,Channel,Filter,,Pivot,BW,lambda_blue,Lambda_red,FWHM (arcsec),Exposure time,flux,err,Lines and dust,,Cont.,HI,H2,Other
NIRCam,Short,F162M,F150W2,1.626,0.168,1.542,1.713,0.053,1933,0.113,0.017,,Cont (81\%); \HI~(10\%); H$_2$ (4\%); \HeI~(1\%),81\%,10\%,4\%,\HeI~(1\%)
,,F212N,,2.121,0.027,2.109,2.134,0.069,1933,0.9,0.08,,H$_2$ (85\%); Cont (13\%),13\%,,85\%,
,Long,F300M,,2.996,0.318,2.831,3.157,0.097,483,0.1511,0.0002,,Cont (58\%); H$_2$ (27\%); \HI~(10\%); \HeII~(1\%),58\%,10\%,27\%,\HeII~(1\%)
,,F335M,,3.365,0.347,3.177,3.537,0.109,483,0.2124,0.0003,PAHs,Cont (53\%); H$_2$ (37\%); \HI~(6\%),53\%,6\%,37\%,
MIRI,,F560W,,5.6,1.2,,,0.207,444,0.44,0.02,,Cont (62\%); H$_2$ (32\%); \HI~(2\%),62\%,2\%,32\%,
,,F770W,,7.7,2.2,,,0.269,444,0.816,0.01,PAHs,Cont (68\%); H$_2$ (14\%); \ArII~(9\%); \HI~(5\%),68\%,5\%,14\%,\ArII~(9\%)
,,F1000W,,10.0,2.0,,,0.328,444,1.86,0.02,,Cont (46\%); \SIV~(26\%); \ArIII~(14\%); H$_2$ (11\%),46\%,,11\%,\SIV~(26\%); \ArIII~(14\%)
,,F1130W,,11.3,0.7,,,0.375,444,0.7253,0.003,PAHs,Cont (94\%); \NiI~(3\%);,94\%,,,\NiI~(3\%)
,,F1280W,,12.8,2.4,,,0.420,444,1.20,0.02,,\NeII~(49\%);  Cont (45\%); \HI~(2\%); H$_2$ (1\%),45\%,2\%,1\%,\NeII~(49\%)
,,F1500W,,15.0,3.0,,,0.488,444,7.43,0.02,,\NeIII~(94\%);  Cont (4\%),4\%,,,\NeIII~(94\%)
,,F1800W,,18.0,3.0,,,0.591,444,4.01,0.02,,\SIII~(54\%); Cont (42\%),42\%,,,\SIII~(54\%)
,,F2100W,,21.0,5.0,,,0.674,444,5.57,0.03,,Cont (71\%); \SIII~(25\%); \ArIII~(1\%),71\%,,,\SIII~(25\%); \ArIII~(1\%)
,,F2550W,,25.5,4.0,,,0.803,444,20,2,,Cont (88\%); \OIV~(8\%),88\%,,,\OIV~(8\%)
MIRI Simul,,F770W,,,,,,,951,,,,,,,,
,,F1130W,,,,,,,951,,,,,,,,
,,F1000W,,,,,,,951,,,,,,,,
\end{filecontents*}
\csvreader[tabular= ll cc lr  r{@}{$\pm$}l rrrlll,
				table head=\hline Instrument  &  Filter  & $\lambda_p$ & BW & PSF  & $t_{\rm exp}$ &  \multicolumn{2}{c}{$F_{\rm tot}$} &  \multicolumn{4}{l}{North Spec contributions} & Notes \\  
                                   & & ($\mu$m) & ($\mu$m) & (\arcsec) & (sec) &  \multicolumn{2}{c}{(Jy)} &  Cont & H~{\sc i} & H$_2$ & Others \\ \hline,
				table foot=\hline ] 
				{observing_log.csv}
				{} 
				{\csvcoli   & \csvcoliii   & \csvcolv  & \csvcolvi & \csvcolix  & \csvcolx   & \csvcolxi  & \csvcolxii  & \csvcolxv & \csvcolxvi  & \csvcolxvii & \csvcolxviii & \csvcolxiii} 
\\
\end{center}
\begin{flushleft}
$\lambda_p$: pivot \citep{Tokunaga2005} wavelength in $\mu$m 
\footnote{\url{https://jwst-docs.stsci.edu/jwst-near-infrared-camera/nircam-instrumentation/nircam-filters}}
\footnote{\url{https://jwst-docs.stsci.edu/display/Latest/MIRI+Filters+and+Dispersers}}. 
BW: band width in $\mu$m. 
PSF: FWHMs of PSFs  in arcsec. For NIRCam simulated values are taken.
$t_{\rm exp}$: exposure time on source in sec.
$F_{\rm tot}$: 
photon-weighted
mean flux density \citep{2014PASP..126..711B} of the nebula in Jy (Sect.\,\ref{sect-total-flux}).
North Spec contributions: Estimated contributions to the images in each filter, calculated using NIRSpec-IFU and MIRI-MRS observations of a region in the northern part of the bright ring (Fig.~\ref{globules-locations}; van Hoof et al. in prep).
Note that some filters are designated to detect specific PAH features, though these PAH bands are not the strongest contributors to the in-band flux in the North spectrum. The PAH spectra observed in the MIRI MRS spectra are discussed by Clark et al. (in prep.)
\end{flushleft}
\end{table*}

\begin{figure*}
    \includegraphics[trim={0.1cm 0.1cm 0.1cm 0.1cm},clip, width=0.75\textwidth]
    {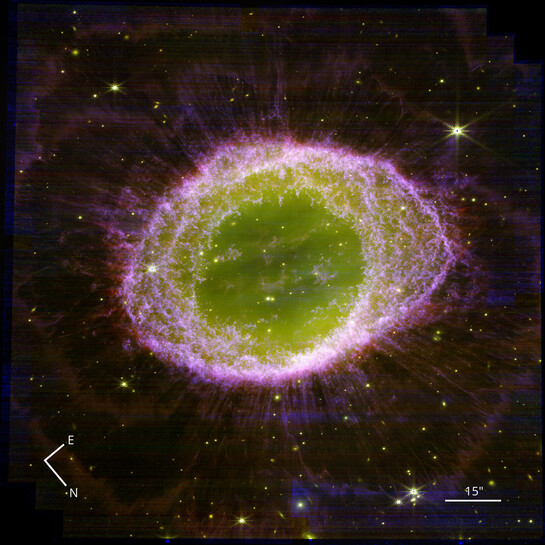}
\caption{NIRCam three colour image: F212N (blue), F300M (green) and F335M (red). This is produced from the pipeline level3 products which does not include the $1/f$ noise removal step. Directions of north and east are indicated
}
  \label{NIRCam-threecolor} 
\end{figure*}
\begin{figure*}
    \includegraphics[trim={0.1cm 0.1cm 0.1cm 0.1cm},clip, width=0.75\textwidth]{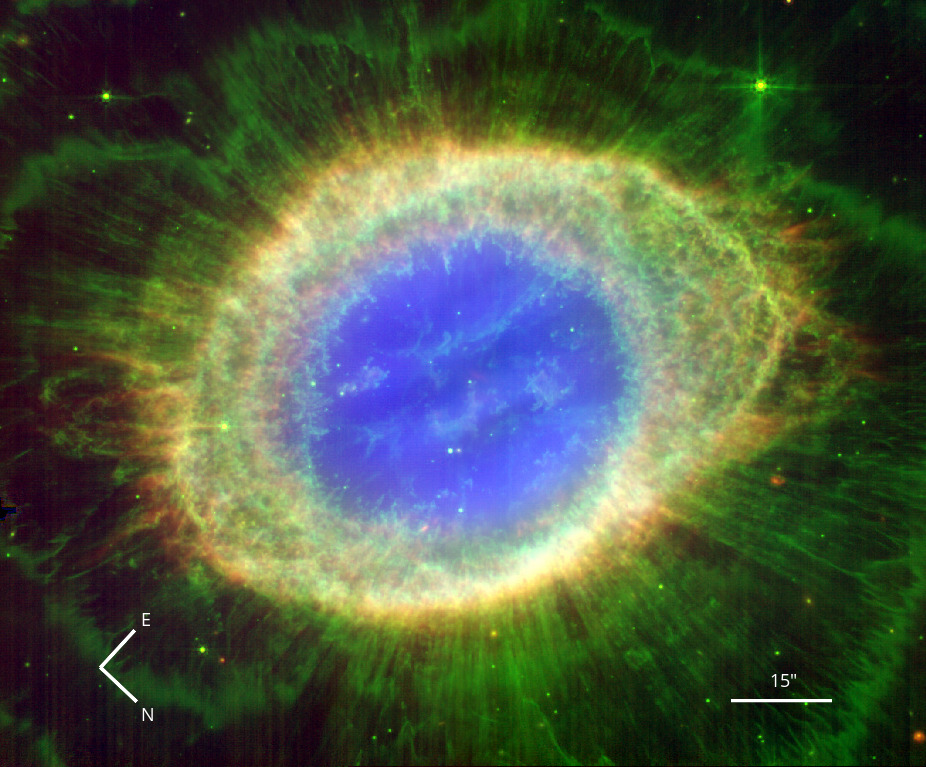}
\caption{MIRI three-colour image: F2550W (blue), F560W (green) and F1130W (red). Directions of north and east are indicated.
}
  \label{MIRI-threecolor} 
\end{figure*}

\section{Basic properties of the Ring Nebula}

The Ring Nebula (NGC~6720 or M~57) is located at a distance of 790$\pm$30~pc, as derived from the \textit{Gaia} parallax of the central star of $1.2696\pm0.0438$ mas \citep{2021A&A...649A...4L}. It has a visual diameter of about 4~arcmin (equivalent to $\sim2\times10^5$~au or $\sim$1~pc) and shows a complex morphology \citep{ODell2013}. The distance  puts the central
star 190~pc above the Galactic plane, consistent with membership of the thin disk. 

\textit{HST} narrow-band images at a variety of optical wavelengths \citep{ODell2004, ODell2007, ODell2013} have been used to derive plasma parameters and extinction maps for the nebula \citep{Ueta2021}. They find an extinction $c_{\rm H\beta}=0.2$, increasing to 0.4 in the shell.\footnote{We use the relation $c_{\rm H\beta}=1.46\,E(B-V)$.} The interstellar foreground extinction contributes $c_{\rm H\beta}=0.1$; the remainder of the extinction is internal to the circumstellar shell.  The extinction is highest in micro-structures (clumps) in the shell \citep{Ueta2021}. 

The star currently has a temperature of 
$T_{\rm eff}=1.35 \times 10^5$~K and a luminosity of 
$L\approx 310~{\rm L}_\odot$ (Sahai et al, in prep.) 
which places it on the white dwarf cooling track. The current mass and progenitor mass of the central star are difficult to determine for objects 
on the cooling track, as the tracks 
tend to converge in that region of the HR diagram. 
\citet{Gonzalez-Santamaria2021} quote 0.58~\Msun\, and  1.5~\Msun, respectively, while the models of \citet{MillerBertolami2016} for the current temperature and luminosity are also consistent with a progenitor mass close to 2\,\Msun.  

A post-AGB star of these masses reaches peak temperature at a luminosity $L\sim 3000\,L_{\odot}$. Fom this point, the luminosity declines to $\sim 200$\,$L_{\odot}$ within a few hundred years, before the decline slows down \citep{MillerBertolami2016}. 
The  stellar luminosity of NGC\,6720 ($L\approx 310~{L}_\odot$) indicates that the star is currently in this phase of rapid fading. Part of the ionized nebula is likely recombining.


\section{Observations and data reduction}

\subsection{JWST Observations}
The Ring Nebula was observed with \textit{JWST} \citep{Gardner2023} in Cycle 1 General Observers (GO) program 1558. The observations were carried out in July and August 2022, using both NIRCam \citep{Rieke2023} and MIRI \citep{Wright2023}.

The NIRCam observations were obtained on 2022 August 4th. Four filters were used, with two filters for each of the Short Wavelength Camera (F162M and F212N) and the Long Wavelength Camera (F300M and F335M).  The nebula was covered by a single field of view (FOV) with a 4-point dither pattern yielding a FOV of 2.45~arcmin with some gaps near the edges of the fields which depend on the camera. The field centre is at 18:53:35.079, +33:01:45.03 (J2000).
The observing log is summarized in Table~\ref{observing_log}.  NIRCam images of individual filter bands are shown in Figure~\ref{NIRCam-individual}.

Imaging observations with MIRI were carried out on 2022 August 20th, using nine filters. The exposure time was 444\,sec in each filter.
The nebula was covered by a 1$\times$2 mosaic with a 4-point dither pattern. The field size was 2.35$\times$1.9~arcmin, again centred at 18:53:35.079, +33:01:45.03 (J2000). 

Additionally, MIRI images in three filters (F770W, F1130W and F1000W) were obtained in simultaneous imaging mode during separate MIRI MRS observations  on 2022 July 4th. These images are centred on 18:53:34.510,
+33:02:09.11, which  allows for some overlap with the direct MIRI images. The simultaneous observations capture part of the outer regions of the Ring Nebula. Their exposure time was 951~sec per filter, hence these simultaneous observations have better sensitivities than those for the main field.

The pixel scales of the reduced data  are 0.031~arcsec per pixel for the NIRCam short wavelength camera images, 0.063~arcsec for the NIRCam long wavelength camera images, and 0.111~arcsec for the MIRI images.

\subsection{Data reduction}

The NIRCam imaging exposures were reduced using the \textit{JWST} Calibration Pipeline\footnote{Available at \url{https://github.com/spacetelescope/jwst}} \citep{Bushouse2022} version 1.9.6 with CRDS version 11.16.19 and CRDS context `jwst\_1075.pmap'.  Each of the NIRCam short wavelength and long wavelength exposures were processed through the \texttt{Detector1Pipeline} with the default parameters.  At this point the ramp slope \texttt{\_rate.fits} images had some of the $1/f$ noise removed using a stand-alone python code provided by Chris Willott outside of the pipeline\footnote{\url{https://github.com/chriswillott/jwst}} before going on to the subsequent data reduction steps.  The files were then processed through the \texttt{Image2Pipeline} and \texttt{Image3Pipeline} stages with the default parameters for all steps, except that the \texttt{tweakreg} step was called with the option to align the images to \textit{Gaia} Data Release 2\footnote{\url{https://www.cosmos.esa.int/web/gaia/dr2}} stars in the field.  The \textit{JWST} pipeline did not have the capability to align to \textit{Gaia} Data Release 3 at the time of the reduction of the images.  The difference in the astrometry between using \textit{Gaia} Data Release 2 or \textit{Gaia} Data Release 3 as the reference is expected to be 10 mas or less, based on comparison of the two catalogues and on a test of this option in pipeline version 1.11.1 on a different data set.  For each of the four filters this resulted in a combined, resampled image that was used for the subsequent analysis.

It was observed that the $1/f$ noise removal interacted with the \texttt{Image3Pipeline skymatch} step to produce a low-level artefact in the short wavelength images, in the regions to the left and right of the main nebula that are observed in 3 of the 9 mosaic positions.  It is not clear why this happens in the \texttt{skymatch} step, since direct examination of the images before and after the $1/f$ noise removal does not show any obvious sign of an offset in the sky background level.  This artefact is seen in the ratio image (see Fig.~\ref{NIRCam-color-2}).

We processed all MIRI imaging exposures using \textit{JWST} Calibration Pipeline version 1.8.3 with CRDS version 11.16.16 and context `jwst\_1062.pmap'. Each of the raw MIRI files was processed through \texttt{Detector1Pipeline} with default parameters. The world coordinate system (WCS) in pipeline-processed \textit{JWST} data can often be incorrect, resulting from uncertainties in the pointing information that are introduced by guide star catalogue errors and roll uncertainty \citep[see][]{pontoppidan2022}. We corrected the WCS reference keywords in the \texttt{Detector1Pipeline} output by determining  and applying the median offset between point sources in preliminary, fully calibrated exposures to their \textit{Gaia} Data Release 3\footnote{\url{https://www.cosmos.esa.int/web/gaia/dr3}} counterparts. We then processed the resulting files through \texttt{Image2Pipeline} to create calibrated dither images across all filters. We created combined mosaics for each MIRI filter, which include the simultaneous imaging field for F770W, F1130W and F1000W, using \texttt{Image3Pipeline}.

MIRI colour images were created by combining multiple filters, where the angular resolutions were matched. For each pair of images, we convolved the images using simulated PSFs, calculated using WebbPSF \citep{Perrin2014}.
 The convolution was processed using the {\sc python} code {\sc PSF Matching}  \citep{Gordon2008, Aniano2011}.

\subsection{Photon-weighted mean flux densities }\label{sect-total-flux}

Table~\ref{observing_log} lists 
the photon-weighted mean flux densities 
 ($F_{\rm tot}$ in Jy) of the main body (the bright shell) of the Ring Nebula. An elliptical aperture was placed centred on the central star. The major and minor radii of the elliptical aperture are 48~arcsec and 38~arcsec respectively, and the rotation angle is 32$\arcdeg$ clockwise from north, aligned with the major axis of the nebula.
The background flux level was estimated from apertures at two areas having the lowest backgrounds within the image (RA=18:53:30.225, Dec=+33:01:55.80, and RA=18:53:40.480, Dec=+33:02:03.91, with a width of 8~arcsec and height of 4~arcsec and 140.52$\arcdeg$  
 as the NIRCam tilt angle). These apertures are shown in Figure~\ref{SED-locations}. The difference between these two background apertures is taken as an estimate of the uncertainty in the total flux. These uncertainties represent a systematic error (background subtraction) and not Poisson noise.
 
The flux in each filter includes emission lines and bands, bound-free and free-free continuum, and dust emission. The important contributions to each filter are listed in Table~\ref{observing_log}, based on NIRSPEC and MRS observations of a region in the northern part of the bright ring (Fig.~\ref{globules-locations}; van Hoof et al. in prep).  The relative contributions will vary across the nebula.
\vspace{3 mm}

\begin{figure*}
    \includegraphics[width=0.85\textwidth]{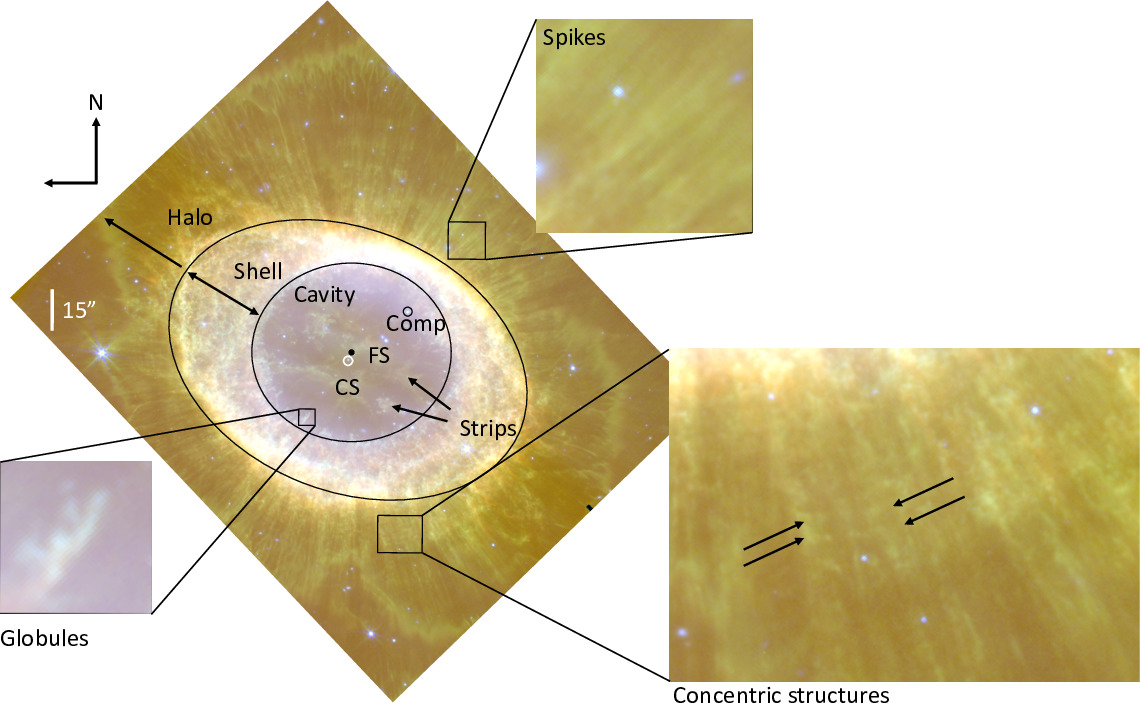}
\caption{The names of the nebula components superposed on the NIRCam and MIRI three-colour image: F300M (blue), F560W (green) and F770W (red). CS (the white circle) is the central star, FS (black spot) is the first-moment centroid of the flux in the F300M image, and Comp is the companion star candidate. North is at the top in this image. The locations of some low-contrast concentric features are indicated, but they are much more easily seen in Fig.~\ref{concentricfeatures}.
}
  \label{components} 
\end{figure*}

\begin{figure*}
  \begin{minipage}[c]{1\textwidth}
    \includegraphics[trim={0.2cm 0.2cm 0.2cm 0.2cm},clip, width=0.5\textwidth]{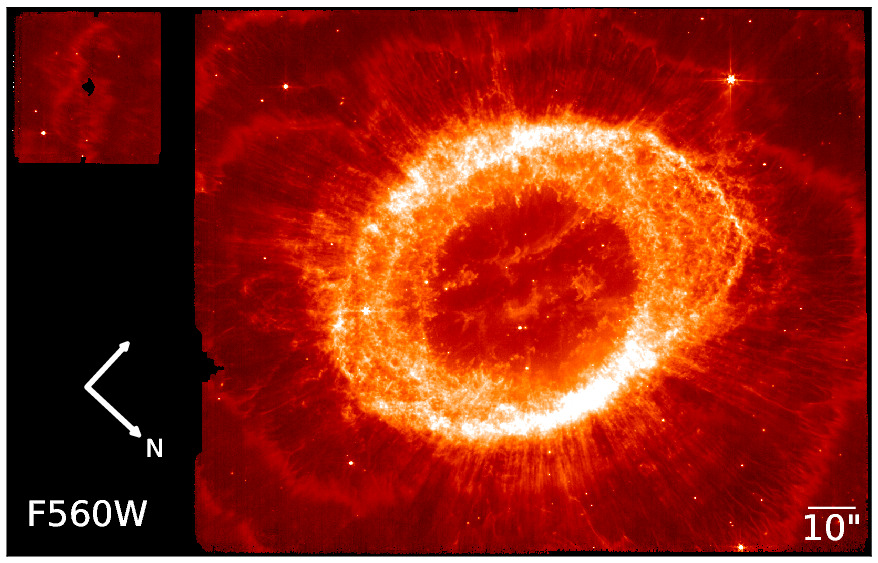}
    \includegraphics[trim={0.2cm 0.2cm 0.2cm 0.2cm},clip, width=0.5\textwidth]{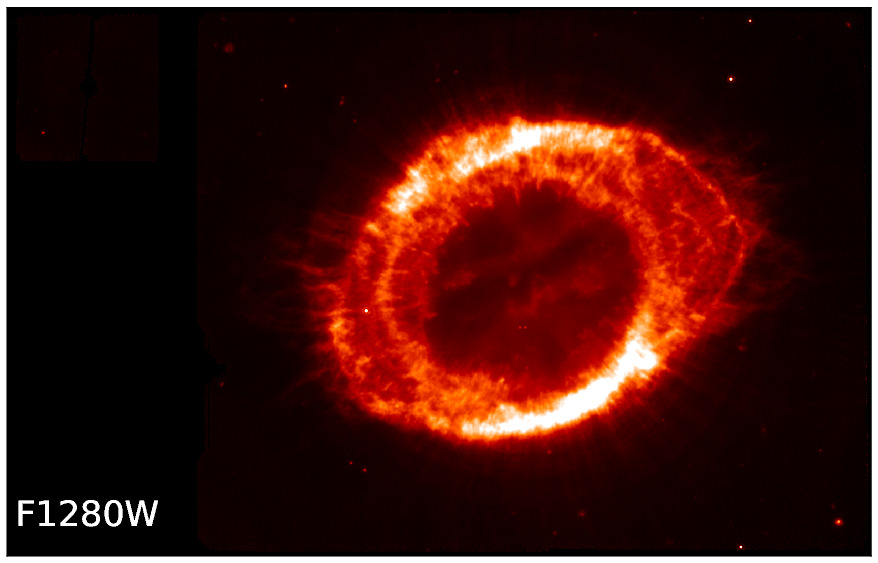}
    \includegraphics[trim={0.2cm 0.2cm 0.2cm 0.2cm},clip, width=0.5\textwidth]{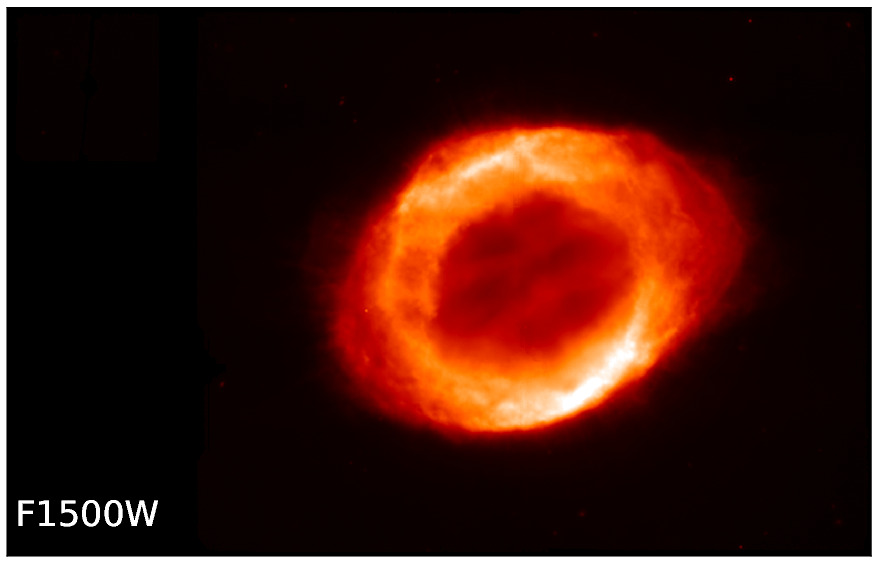}
    \includegraphics[trim={0.2cm 0.2cm 0.2cm 0.2cm},clip, width=0.5\textwidth]{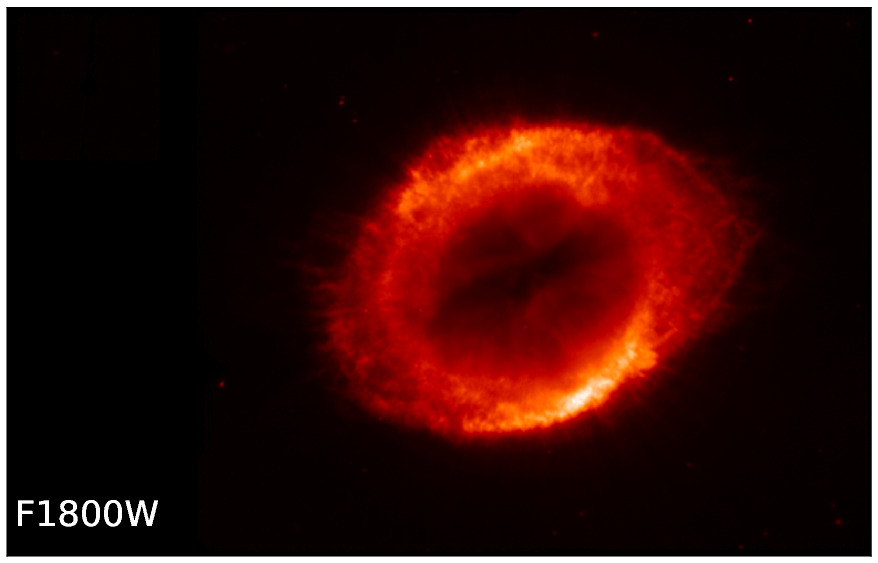}
    \includegraphics[trim={0.2cm 0.2cm 0.2cm 0.2cm},clip, width=0.5\textwidth]{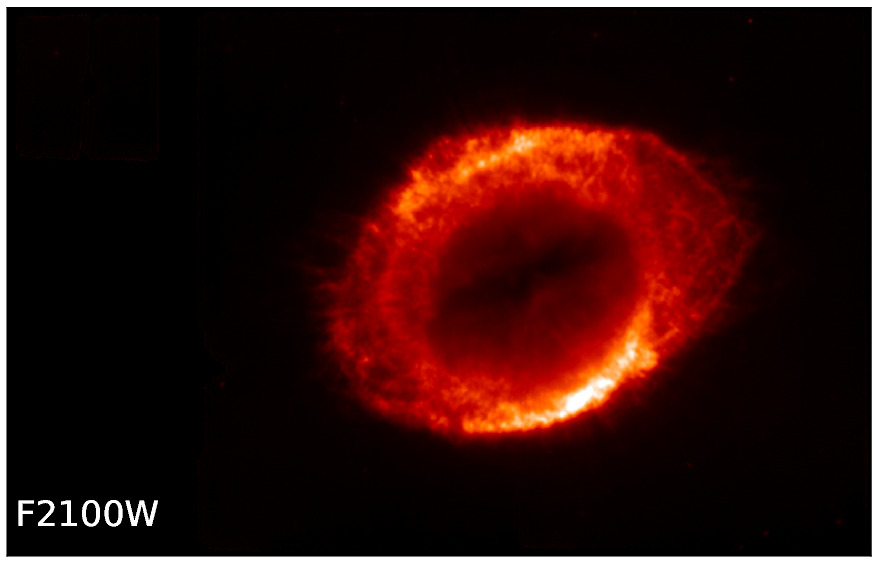}
    \includegraphics[trim={0.2cm 0.2cm 0.2cm 0.2cm},clip, width=0.5\textwidth]{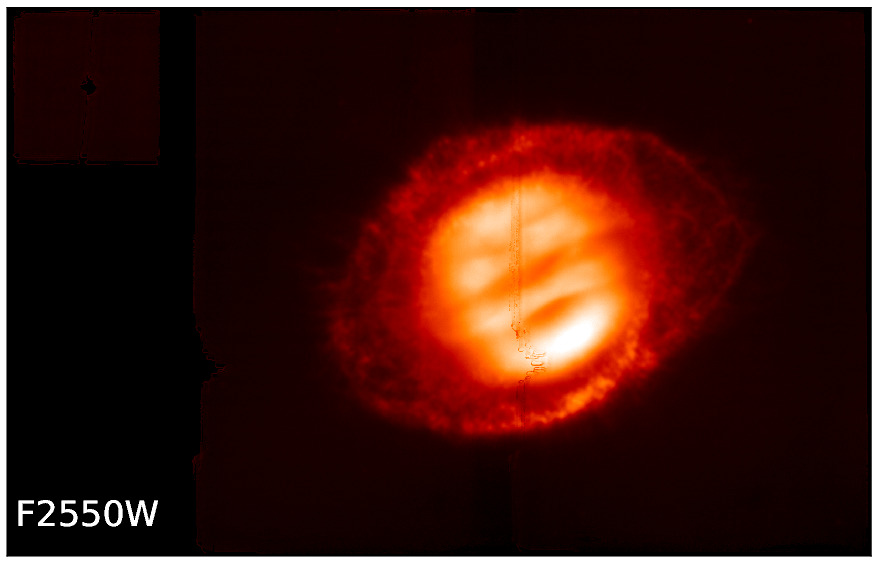}
      \end{minipage}\hfill
\caption{ MIRI images of the Ring Nebula.
}
  \label{MIRI} 
\end{figure*}

\begin{figure*}
  \begin{minipage}[c]{1\textwidth}
    \includegraphics[trim={0.2cm 0.2cm 0.2cm 0.2cm},clip, width=0.5\textwidth]{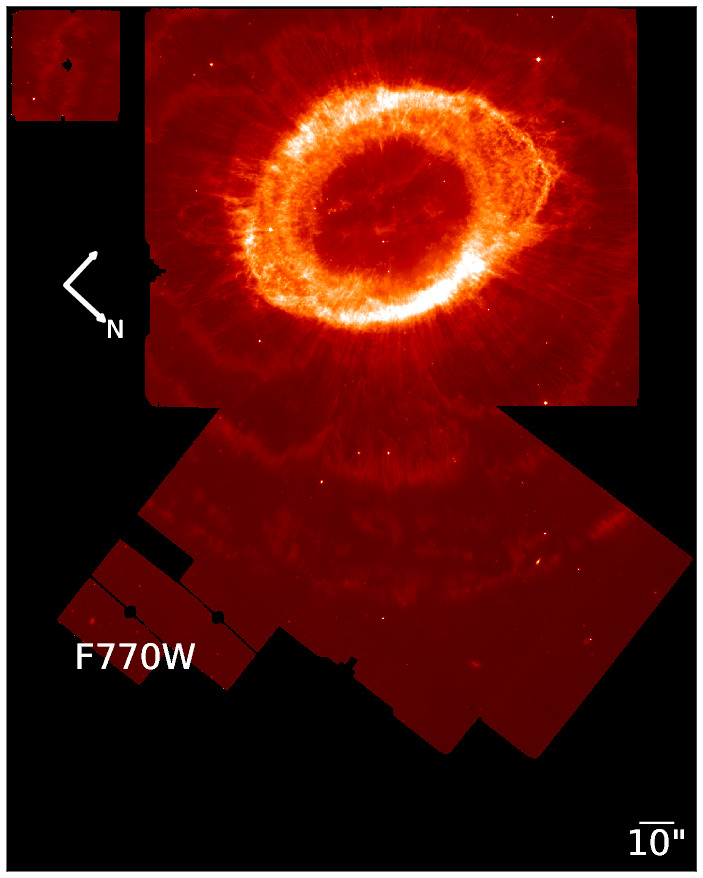}
    \includegraphics[trim={0.2cm 0.2cm 0.2cm 0.2cm},clip, width=0.5\textwidth]{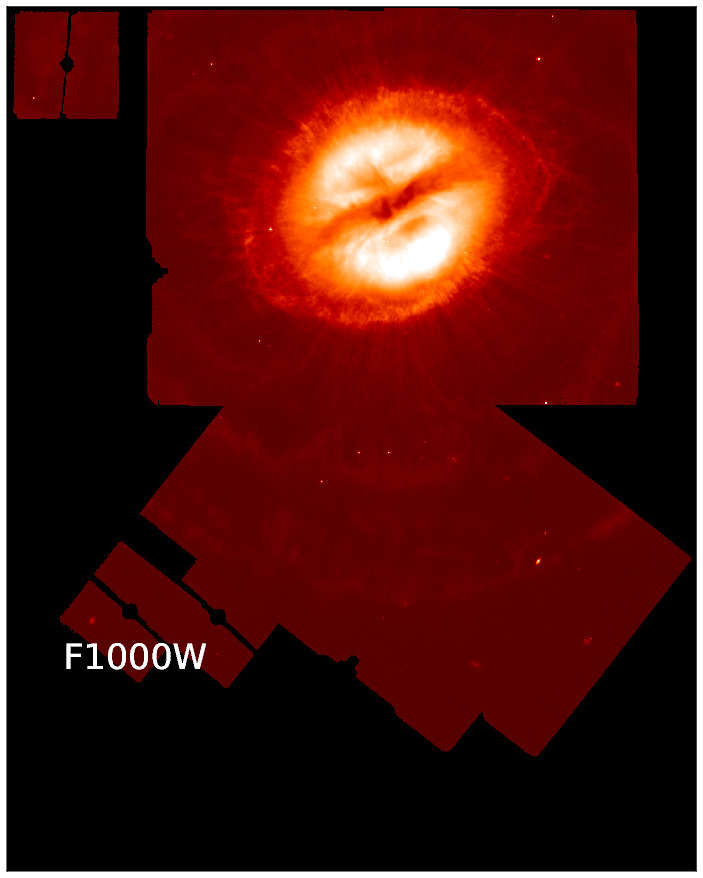}
    \includegraphics[trim={0.2cm 0.2cm 0.2cm 0.2cm},clip, width=0.5\textwidth]{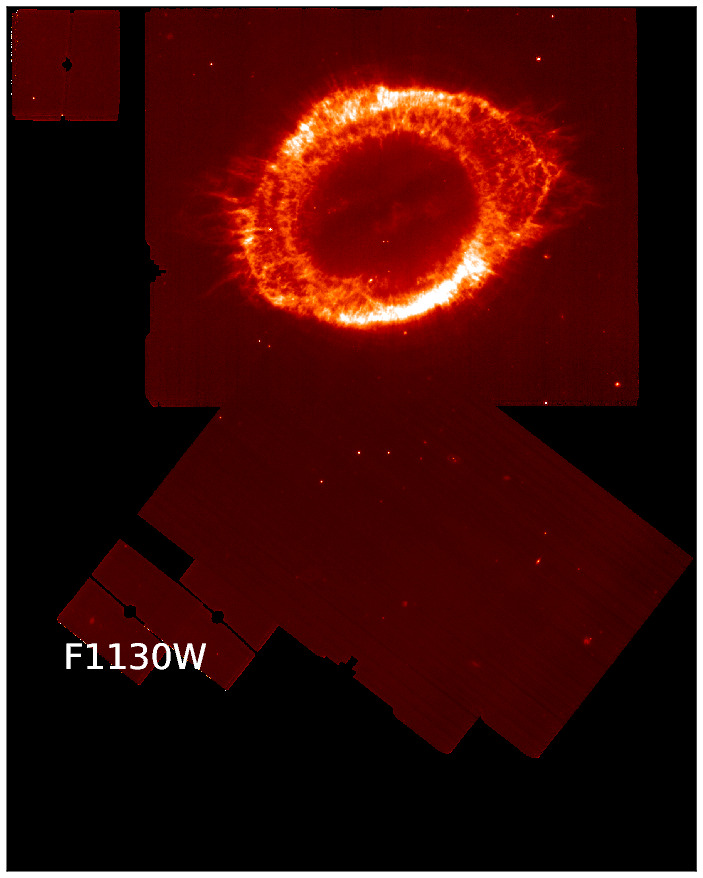}
      \end{minipage}\hfill
\caption{ 
MIRI images combining the direct mapping images and the simultaneous images of the Ring Nebula which extend the imaged field further to the north and west.
  \label{MIRI-simultaneous} }
\end{figure*}

\section{Description and analysis of the images}

\subsection{Stratification}

The nebula can be divided into three clearly distinct regions: the central cavity, the bright shell and the halo.  \citep[The latter is sometimes divided into an inner and an outer halo;][]{Balick1992}. This structure can be recognized in the multi-colour images of  Figure~\ref{NIRCam-threecolor} (NIRCam) and Figure~\ref{MIRI-threecolor}  (MIRI). 
The components are labelled in Figure~\ref{components}.  In these images, the regions show different colours. This indicates that they differ in their line emission: the nebula is stratified.

The low-density central cavity appears as an approximately circular structure, with a radius of about 25~arcsec, which emits mainly in the F1000W  and F2550W filters (Figs.~\ref{MIRI} and ~\ref{MIRI-simultaneous}).
These two filters contain the high excitation \SIV\, (ionization potential of 35~eV) and \OIV\, (55~eV) lines (Table~\ref{observing_log}). 
All other filters show the cavity as a low-emission region (Figs.~\ref{NIRCam-threecolor}, \ref{MIRI-threecolor}, \ref{MIRI} and \ref{NIRCam-individual}). The cavity thus has a higher excitation than the surrounding nebula. 

The cavity contains a linear structure approximately along the long axis of the nebula, which is visible in most filters and consists of two brighter stripes on either side of the central star. The region between these stripes shows little emission in all filters except for F2550W. The long wavelength F1800W and F2100W filters do not show the stripes. A somewhat similar structure is present in the central region of NGC\,3132 \citep{DeMarco2022}.  \citet{ODell2013}  interpret the stripes as features in the inner halo, seen in projection against the cavity. 

The shell surrounding the cavity is a broad region with a well-defined inner and outer edge. It is bright in all filters. Unlike the central region, it has an elliptical shape. The outer radius is 44~arcsec along the major axis and 35~arcsec along the minor axis. The position angle of the major axis is 132\arcdeg\ (from north to west). 
The shell is significantly brighter along the minor axis than the major axis. 

The outer edge of the shell is somewhat distorted towards the northeast. This is approximately the direction of the \textit{Gaia} DR3 proper motion of the central star (NNE, 10\,km\,s$^{-1}$), and it is possible that this distortion is related to interaction with the interstellar medium.

The emission in the shell is clumpy, especially in filters dominated by H$_2$ (Table~\ref{observing_log}). 
The H$_2$ emission from the shell has long been known
\citep{Greenhouse1988}, and was known to be clumpy \citep{Speck2003}, but is seen at much higher angular resolution and sensitivity with \textit{JWST}. Close inspection shows evidence for a very large number of clumps or globules, seen throughout the shell. The almost complete lack of clumps seen projected on the central cavity supports the interpretation of the shell as an equatorial or toroidal structure, seen approximately pole-on \citep{Bryce1994, ODell2013}.

The inner halo is seen in all NIRCam images, apart from F162M (Fig.~\ref{NIRCam-individual}). The inner halo is also seen in the MIRI F560W and F770W MIRI images; the filters that clearly show the halo are dominated by H$_2$ lines. H$_2$ emission from the halo was detected by \citet{vanHoof2010}. The faint halo shows a wealth of structure, including concentric arcs, radial stripes and distorted bright edges. These are each discussed separately in the following sections. The simultaneous fields show a part of the outer halo, terminating at a radius of 1.9~arcmin from the centre. The (incomplete) coverage is consistent with a generally circular outer halo. 

The central star is clearly detected in the NIRCam images, and in the MIRI F560W and F770W images.
The central star and its vicinity will be discussed in a separate paper (Sahai et\,al., in preparation).

\subsection{Offset of the nebular centre from the stellar position}

\begin{figure*}
    \includegraphics[width=1.0\textwidth]{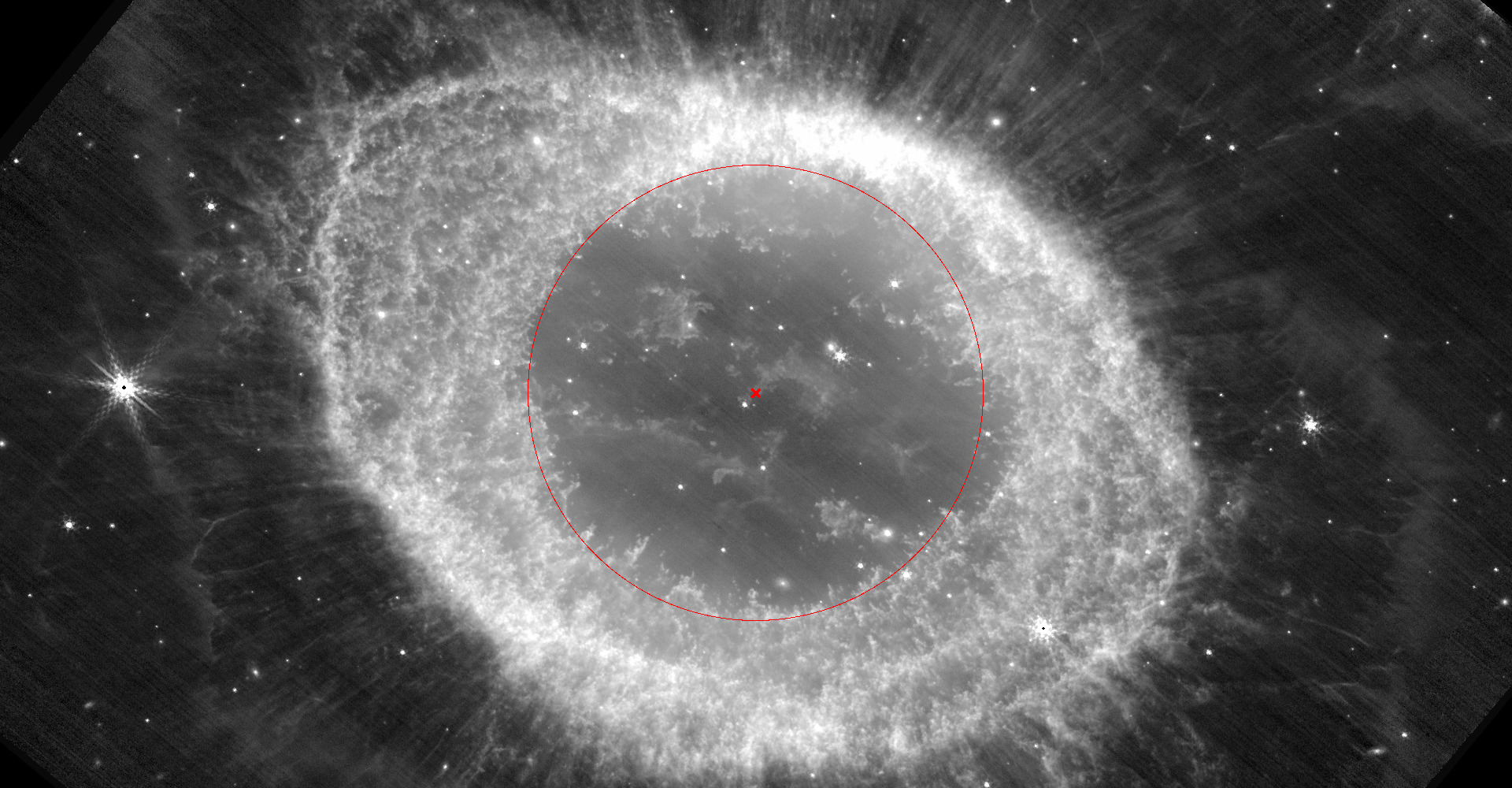}
    \caption{The boundary of the inner cavity of the Ring Nebula is close to circular, but its centre (marked by an X) is offset by 2~arcsec from the central star. The image is in the F335M filter. North is at the top, east is left.\label{innerring}}
\end{figure*}

Although the outer edge of the inner cavity is close to circular, its centre does not exactly correspond to the position of the central star, but instead appears to be offset to the north-west (roughly in the direction of the \textit{Gaia} companion; see Sect. \ref{sec:multi}) 
by about 2~arcsec (Fig.~\ref{innerring}). The flux-weighted centres of the F300M and F335M images (calculated as the first moment of the images) are also offset from the central star, and nearly coincide with the centre of the inner cavity. The coordinates of the flux-weighted centre for the F300M image are RA=18:53:35.04 and Dec=+33:01:46.8, while the coordinates of the centre of the circular inner edge of the cavity are RA=18:53:35.01, Dec=+33:01:46.01.

This offset with the shell and cavity must be considered tentative, as the complex small-scale structures introduce uncertainties. The interpretation is also open to discussion. The structures here are affected by the original mass loss, the ionization and the hot stellar wind. If the offset is interpreted as caused by the original mass loss, for an age of 4000 yr \citep{ODell2013b}, the offset would correspond to a velocity difference between star and nebula of around 2~km\,s$^{-1}$. 

Off-centre central stars are known in some other planetary nebulae. The PN A39 has an offset of 2~arcsec in a symmetric nebula \citep{Jacoby2001}. Another example is Hu2-1 \citep{Miranda2001}.

\subsection{Globules}
\label{clumpfindingtext}

\begin{figure*}
  \begin{minipage}[c]{1\textwidth}
    \includegraphics[trim={0.2cm 0.2cm 0.2cm 0.2cm},clip, width=1.0\textwidth]{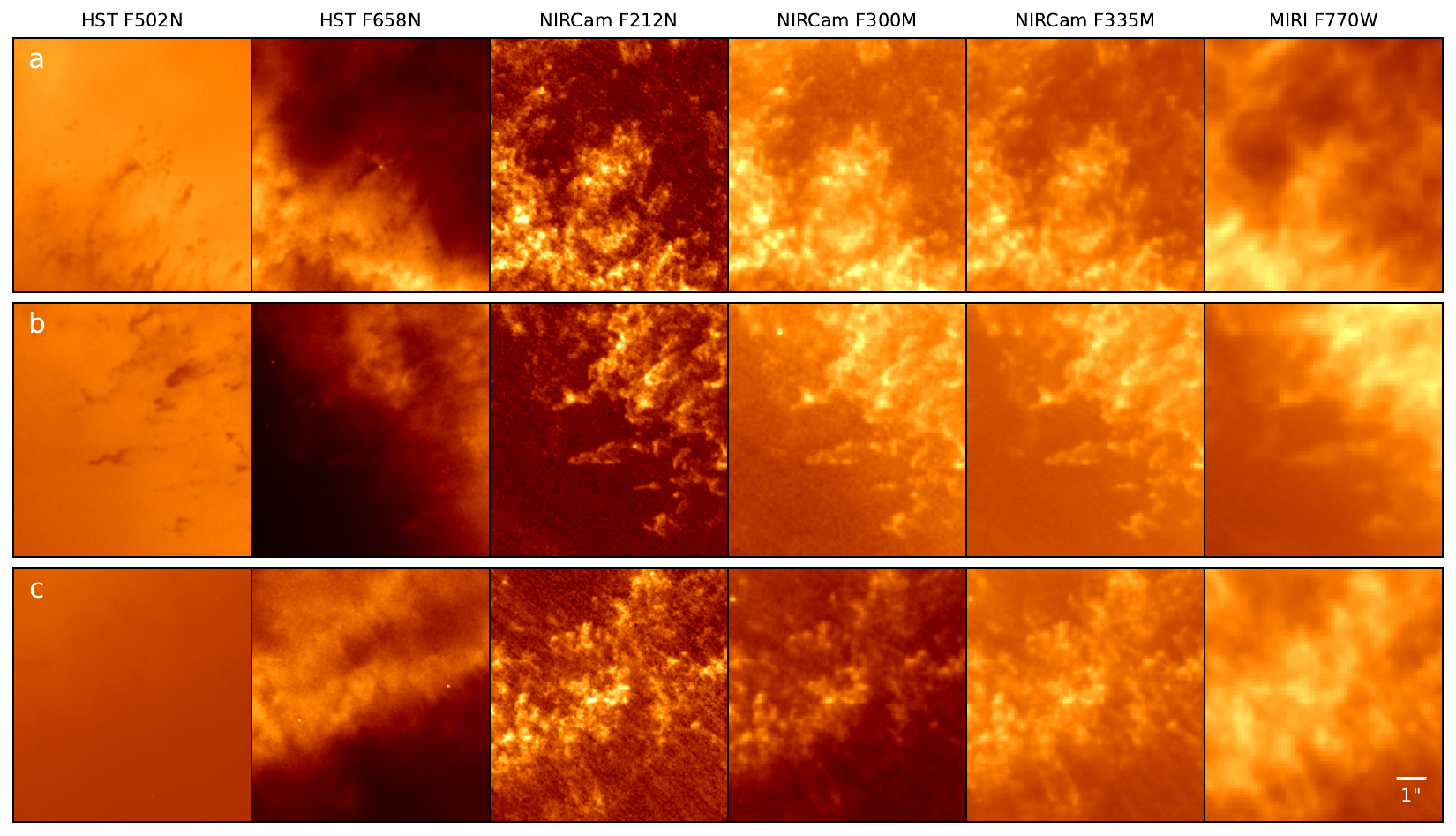}
      \end{minipage}\hfill
\caption{ Zoomed-in images of globules in the Ring Nebula. The locations of the three zoomed-in regions are indicated in Fig.~\ref{globules-locations}. Globules are detected in H$_2$ emission in the \textit{JWST} NIRCam images and MIRI images, while some of them are seen in absorption against the diffuse ionized emission in the \textit{HST} F502N and F658N images.
}
  \label{globules} 
\end{figure*}

Figure~\ref{globules} shows a sequence of images of regions containing globules, covering optical (\textit{HST}) and near- and mid-infrared (\textit{JWST}) wavelengths. The \textit{HST} images show the globules as extinction peaks \citep[e.g.,][]{ODell2013}, while the \textit{JWST} images do not. Even the shortest wavelength NIRCam image at 1.62~$\mu$m shows no absorption (Fig.~\ref{NIRCam-individual}). This is in contrast to NGC\,3132, in which  \citet{DeMarco2022} found some globules showing absorption at 1.87~$\mu$m. We note that the sensitivity to absorption is less in our data because of the choice of filters: the F187N filter used for NGC\,3132 contains a strong HI Pa$\alpha$ line, while the F162M and F212N filter do not contain strong atomic emission lines (Table \ref{observing_log}). There is less absorbable nebular background in our images.  

To estimate the mass of the globules in NGC\,3132, \citet{DeMarco2022} assumed $A_V$=1.7~mag and 3.9~mag for two specific knots. We follow their analysis. 
A direct measurement of the most heavily extinguished globule in Figure~\ref{globules}b gave an extinction $A_{{\rm 502nm}}$=1.1~mag. Using the wavelength dependence of the extinction curve from \citet{1989ApJ...345..245C}, and $A(V)/N(\rm{H})\sim2.3\times10^{21}$\,cm$^{-2}$, the column density becomes $N_{\rm H}= 2.2 \times 10^{21}$\,cm$^{-2}$. The diameter of this globule is about 0.4~arcsec. This yields a density  $n_{\rm H} \approx 5 \times 10^5$~cm$^{-3}$ and a mass of $m \approx 2 \times 10^{-5}\,$\Msun.

The majority of clumps are of order 0.2~arcsec in diameter, or around 150~au. Assuming that the clumps are roughly spherical and have similar density, the mass of such a clump becomes $m\approx 5\times 10^{-6}$~M$_\odot$.

At these densities (noting that $n_{\rm H_2}=0.5 n_{\rm H}$) the globules can be in pressure equilibrium with the ionized gas. The density in the ionized gas is approximately $1.3 \times 10^3$\,cm$^{-3}$ and the temperature is around 9000\,K \citep{Ueta2021}. Pressure equilibrium would be reached for a globule temperature and density $T \approx 200 \times 10^6/n_{\rm H}$\,K. The globules could therefore be essentially stable and avoid collapse or dissipation, until the surrounding gas recombines. Whether they are indeed in pressure equilibrium or are still collapsing depends on the unknown temperature, turbulence and details for the formation process. 

\begin{figure}
    \includegraphics[width=0.48\textwidth]{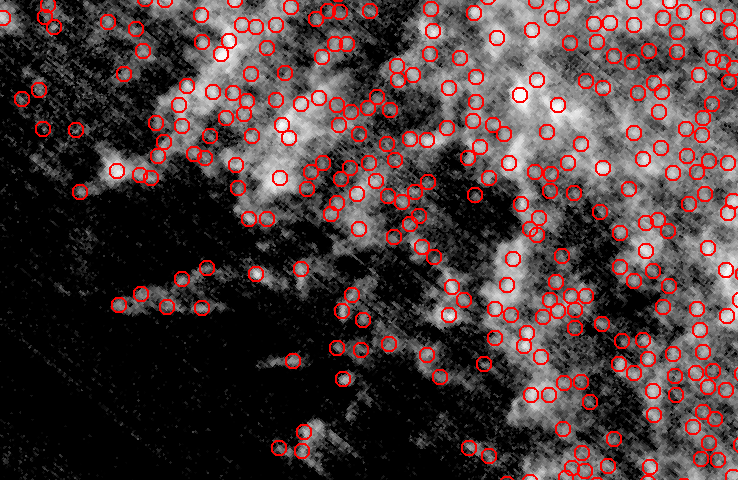}
    \caption{The positions of clumps, shown in red, identified by a peak-finding algorithm in a section of the main ring in the F212N image.}
    \label{clumpfinding}
\end{figure}

Even the angular resolution of the \textit{JWST} images does not allow the system of globules to be cleanly resolved. However, to estimate their total number, we applied a peak-finding algorithm from the Python package {\sc photutils} (\citealt{larry_bradley_2023_7946442}) to the F212N image. A small section of the image is shown in Figure~\ref{clumpfinding}, with the locations of the peaks found by the algorithm indicated. The peak-finding algorithm identifies about 17\,500 peaks, which, given the density of clumps and resulting overlap, is likely to be an underestimate of the total number. A manual count was done in small regions, which, extrapolated to the full area, gives an estimated population of $\sim$25\,000 globules.

Based on these numbers, the clumps have a combined mass of up to 0.1\,M$_\odot$. In comparison, the CO emission also traces $\sim 0.1\,$M$_\odot$ \citep{Bachiller1989}. The molecular mass is similar to that found for NGC\,3132 \citep{DeMarco2022}.

The typical filling factor of dense clumps in planetary nebulae has been estimated as $7 \times 10^{-5}$ \citep{Zhang2004}. In the Ring Nebula, this factor appears considerably higher: the clump filling factor for the shell is $\sim 2 \times 10^{-3}$. The density of the ionized gas in the shell is around $1.3\times 10^3$\,cm$^{-3}$ \citep{Ueta2021}. This gives an ionized mass in the shell of roughly 0.15\,\Msun. The clumps may therefore account for up to half the mass of the shell. \citet{Ueta2021} argue for a similar ratio. 

\citet{Sahai2012} used [C~{\sc ii}] observations made with the \textit{Stratospheric Observatory for Infrared Astronomy} (\textit{SOFIA}) to show that 0.11\,\Msun, half the mass of the nebula, lies in a photon-dominated region (PDR) zone. \citet{Liu2001} found densities of $n_{\rm H}\sim10^5$\,cm$^{-3}$ for these PDR regions. The PDR emission zone is likely associated with the clumps. The PDR zone is mixed in with the ionized gas, in a region where clumps and ionized gas co-exist, rather than forming a separate outer shell. This zonal mixing is also seen in the ionized gas \citep[e.g.,][]{Garnett2001}.

The H$_2$ images reveal few cometary tails emanating from the globules. This is in contrast to the Helix Nebula where the majority of the globules (at least in the inner region)
have well-developed tails seen in H$_2$  \citep{Matsuura2009}. Some short extensions of around 1~arcsec in length can be seen in absorption in the \textit{HST} F502N and F658N images (Fig.~\ref{globules}a,b) but these do not have clear counterparts in the \textit{JWST} images. An exception is the largest globule in Fig.~\ref{globules}b, which shows a faint bow shape in the \textit{HST}/F658N image that becomes a longer `U' shape in the \textit{JWST}/F300M and F335M images. The extensions seen in the \textit{HST} images tend not to be straight. They may trace an early stage of tail formation.

\subsection{Radial spikes}

The halo shows multiple narrow, radial features pointing away from the central star, which following \citet{DeMarco2022} we call `spikes' (Figs.\,\ref{concentricfeatures} and \ref{edgedetection}). \footnote{\citet{ODell2013} uses the word `rays'.} They are seen only outside the bright shell, and mainly in H$_2$.  The spikes are mainly a feature of the inner halo \citep{ODell2013}.

In a section of the F770W image covering 20 degrees of azimuth, we count 15--20 spikes. From this, we estimate that there are about 300--400 spikes in total. The exact number is imprecise due to the low contrast, small separation, and partial overlap of many spikes. The typical width appears to be around 0.4~arcsec. The typical length of the visible spikes is around 20~arcsec, but they may in a few cases extend twice as far, out to the outer edge of the nebula.

The spikes are expected to arise from illumination effects where stellar light escapes through holes in the shell \citep{DeMarco2022}. They line up better with the central star than with the offset centre of emission, which indicates that the dominant cause lies in the current or recent radiation from the star. There are some cases of misalignment with the star, possibly where there is partial overlap between spikes. 

The number of spikes is of order 2~per~cent of the number of globules. This suggests that there is no direct relation between the globules and spikes. \citet{ODell2013} argue that some of the spikes can be the shadows of large  globules but this is not evident from the data here.

\subsection{Arcs -- concentric structures}

\begin{figure*}
\includegraphics[width=\textwidth]{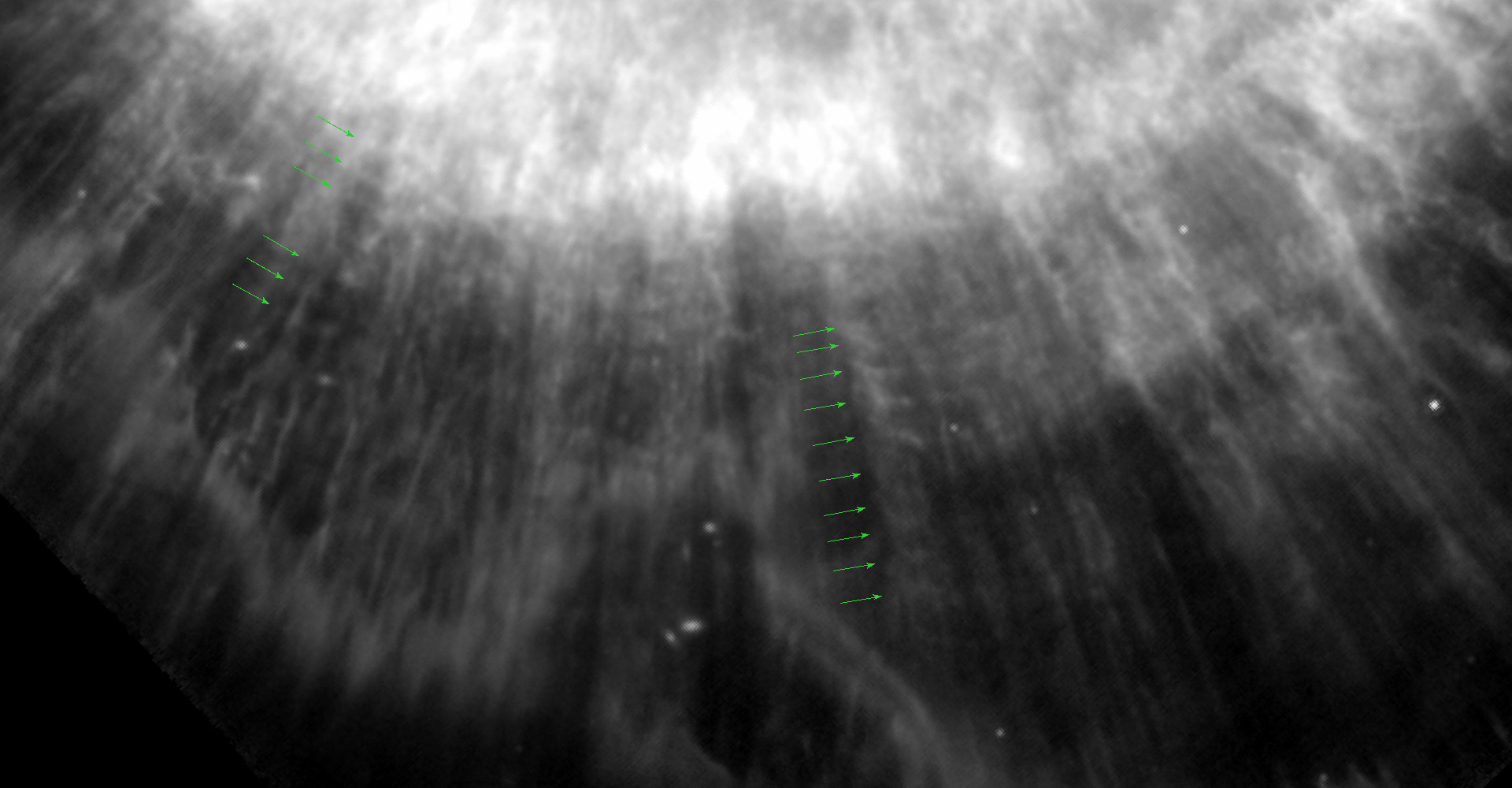}
\caption{Regular concentric features in the outer regions of the F770W image of the Ring Nebula. Green arrows indicate the locations where these regularly-spaced features are most easily seen.}
\label{concentricfeatures}
\end{figure*}

\begin{figure*}
    \includegraphics[width=\textwidth]{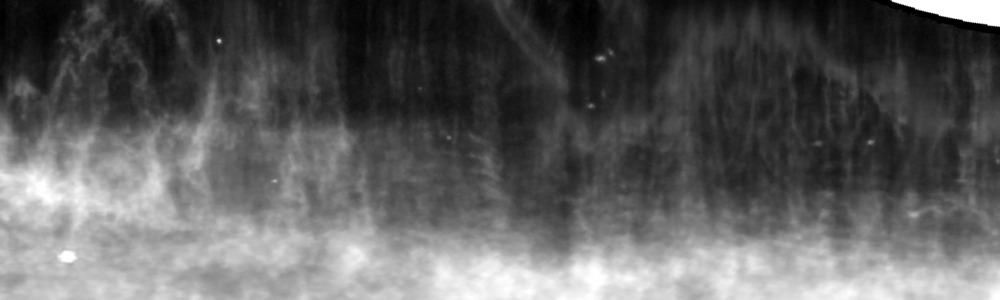}
    \includegraphics[width=\textwidth]{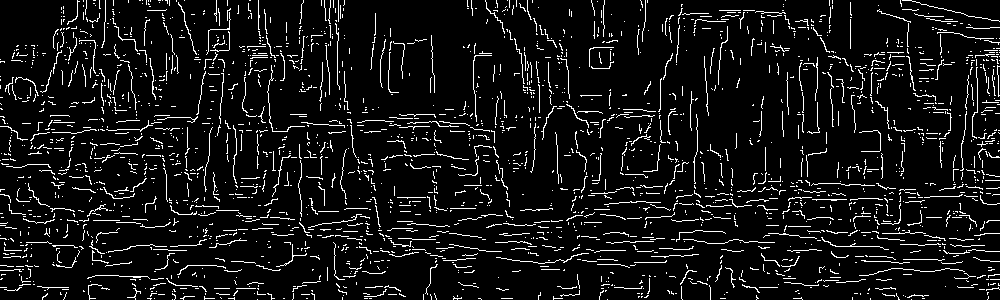}
    \includegraphics[width=\textwidth]{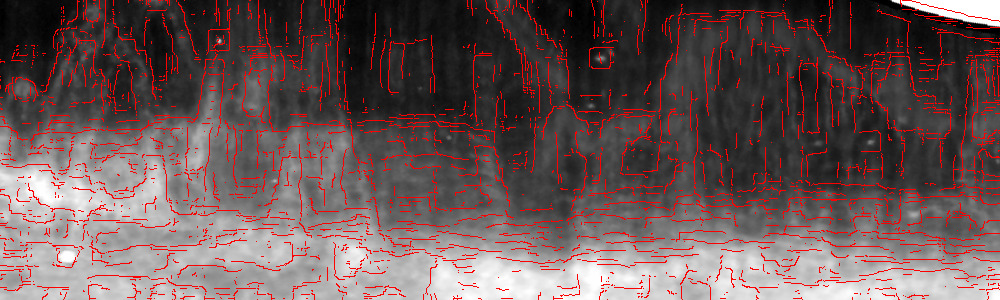}
    \caption{Application of an edge-detection algorithm to a section of the polar-projected image, highlighting the predominantly radial and concentric nature of the nebular structures. Top: reprojected F770W image. Centre: edges detected using the algorithm of \citet{canny86}. Bottom: edges overlaid on the reprojected image. The horizontal direction is azimuthal direction of the nebula, and vertical direction is radial direction, with the central star towards the bottom.  The horizontal axis covers 90\,degrees and the vertical axis extends 45 arcsec.
    \label{edgedetection}}
\end{figure*}

A series of faint, broken concentric arcs is apparent outside the bright ring in the H$_2$ halo. In places, up to 10 arcs can be identified (Figure~\ref{concentricfeatures}). In some directions, the arcs appear to have been disrupted, possibly due to density variations in the nebula, but in general the curvature is regular (Fig.~\ref{edgedetection}).

These concentric structures are most apparent in the F770W image, which contains emission lines of H$_2$, \ion{H}{i} and [\ion{Ar}{ii}]. They are also clearly visible in the F1000W and F1130W images and can be distinguished at longer wavelengths in the F1800W and F2100W filters, but at lower contrast due to the poorer spatial resolution. At shorter wavelengths with better spatial resolution, the much clumpier appearance of the nebula in the lines isolated by these filters makes the features harder to trace. The arcs are seen in most directions but are obscured by the shell along the major axis.

Using a portion of the F770W image in which the arcs are most clearly seen, we measure an average separation of about 1.5~arcsec. Assuming that the arcs are embedded in the outflow and are in the plane of the sky, this separation and outflow velocity provide a time scale. The outflow velocity of the Ring Nebula decreases outward from $\approx40$~\kms~in the cavity to 20--25~\kms~in the shell and 15~\kms~in the halo \citep{ODell2013, Martin2016, Sahai2012}. We assume a value of $20\pm5$~\kms~at the location of the arcs, just outside the shell. At a distance of 790\,pc, this separation corresponds to a time interval of $280\pm70$\,yr. 
This is much too short an interval to be related to periodic thermal pulses that may enhance mass loss, which are expected to occur at intervals of 10$^{4}$--10$^{5}$ years. 

Instead, a common scenario invoked to explain arc systems like these is one in which a close binary companion modulates the outflow from the AGB star. The time interval of 280\,years then corresponds to the orbital period of the companion. If the combined mass of the original AGB star and its companion is $1.5\pm0.5$M$_\odot$, the orbital separation would be about $50\pm15$~au.

\textit{JWST} images of the Southern Ring Nebula (NGC\,3132) have revealed very similar concentric structures, with a separation of about 2~arcsec at a distance of 750~pc. These have also been attributed to the effects of a binary companion, with an orbital period of 290--480~yr \citep{DeMarco2022}.

To highlight the radial and concentric structures, we reprojected the MIRI F770W image into polar coordinates relative to the central star. We then applied the edge-detection algorithm of \citet{canny86} to the reprojected image. The result is shown in Figure~\ref{edgedetection}. Within the shell, the edges are primarily azimuthal, i.e. horizontal in the azimuth--radius plot. Further out, the edges abruptly change to radial (i.e. vertical). This confirms that the spikes start suddenly at the outer radius of the shell. There is a region of overlap between the horizontal arc lines and the vertical spikes. 

\section{Discussion}

\subsection{PAHs}

The polycyclic aromatic hydrocarbon (PAH) feature at 11.3\,$\mu$m has been detected in {\it Spitzer} spectra of the Ring Nebula \citep{Cox2016}. {\it Spitzer}'s Infrared Spectrograph (IRS) covered a 99$\times$18~arcsec area along the major axis of the elliptically-shaped Ring Nebula, with a pixel scale of 1.8~arcsec.\footnote{Program 40536, P.I. H. Dinerstein} This spectral map showed that the PAH feature is emitted from the outer half of the shell, a more or less similar emitting region to the H$_2$ S(1) and S(2) lines at 17.04 and 12.28\,$\mu$m.

\begin{figure*}
        \includegraphics[trim={0cm 0cm 0cm 0cm},clip, width=1.0\textwidth]{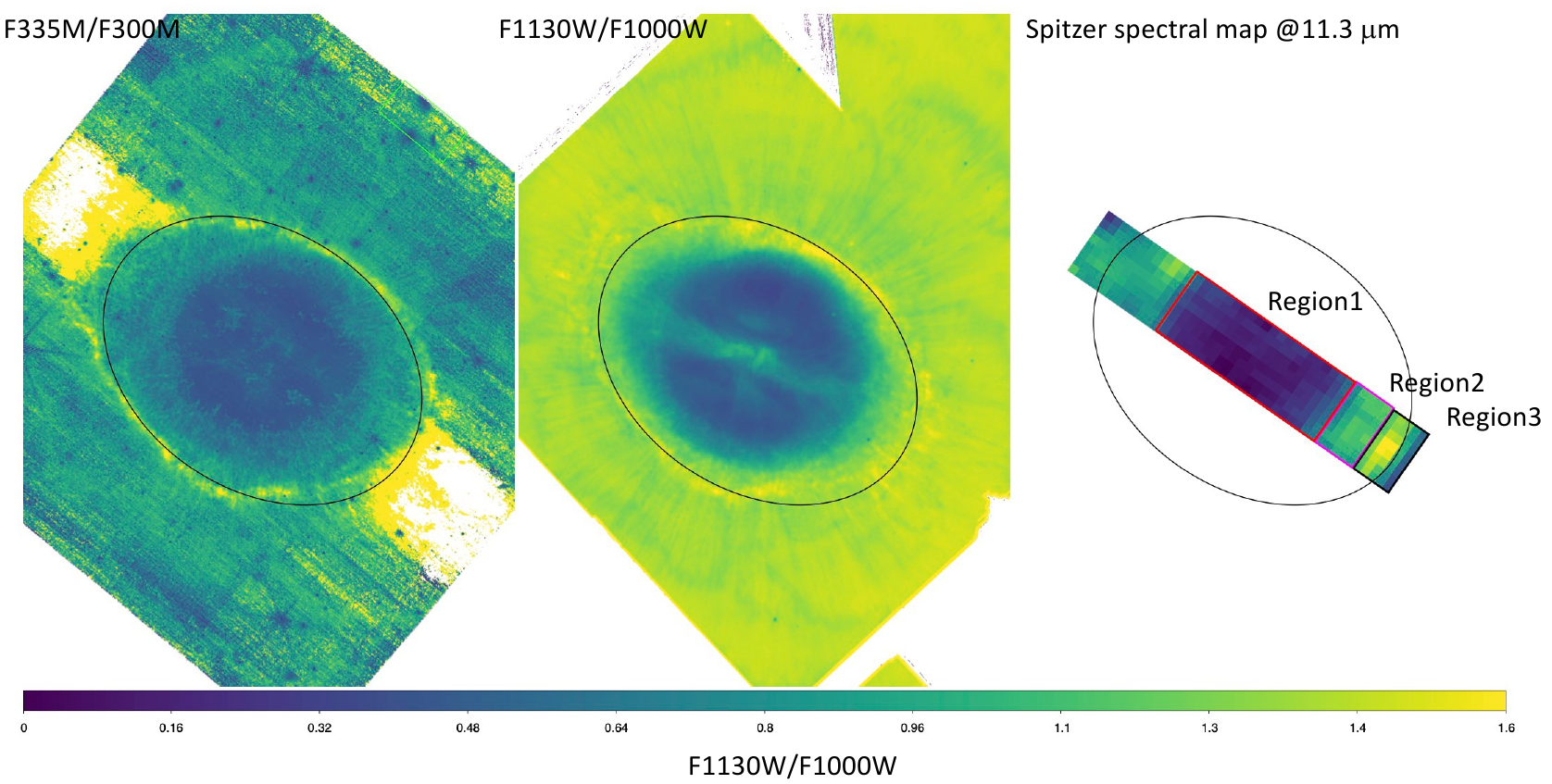}
\caption{ F335M/F300M ratio (left), F1130W/F1000W ratio (middle) and \textit{Spitzer} spectral map at 11.3\,$\mu$m (10.95--11.65\,$\mu$m, corresponding to \textit{JWST} F1130W filter) \citep{Cox2016} on the same scale. The  narrow ring at the edge of the shell, guided by a black ellipse, is interpreted as the location of PAH emission. The \textit{Spitzer} spectra were extracted at three different regions and plotted at Fig.~\ref{SEDs}. 
North is at the top.
}
  \label{NIRCam-color-2} 
\end{figure*}

The higher angular resolution \textit{JWST} images allow the PAH-emitting regions to be studied in more detail. The F335M filter encompasses the 3.3~$\mu$m PAH feature, while the F300M filter lacks this band but otherwise contains similar nebular emission (Table~\ref{observing_log}).  The F1130W filter has the lowest contribution from emission lines (Table~\ref{observing_log}), and best reflects the contributions of the continuum and PAH bands.  Both the F335M/F300M ratio image  and the F1130W/F1000W ratio image (Fig.~\ref{NIRCam-color-2}) show indications of a narrow ring of excess emission located at the outer edge of the shell.

From Table \ref{observing_log}, the PAH contribution to F335M and F1000W is $<14$~per~cent\ and $<7$~per~cent\ (but note that this is estimated from MIRI IFU spectra which are not centred on the ring; Fig.~\ref{globules-locations}).  This ring shows up only in the two filters containing PAH bands (Fig.~\ref{NIRCam-color-2}) and is not seen in other filter combinations (Figs.~\ref{NIRCam-color-1} and \ref{MIRI-color}).  We interpret these narrow ring excesses as possible PAH emission. 

\begin{figure}
     \includegraphics[trim={0.0cm 0.cm 0cm 0.cm},clip, width=0.48\textwidth]{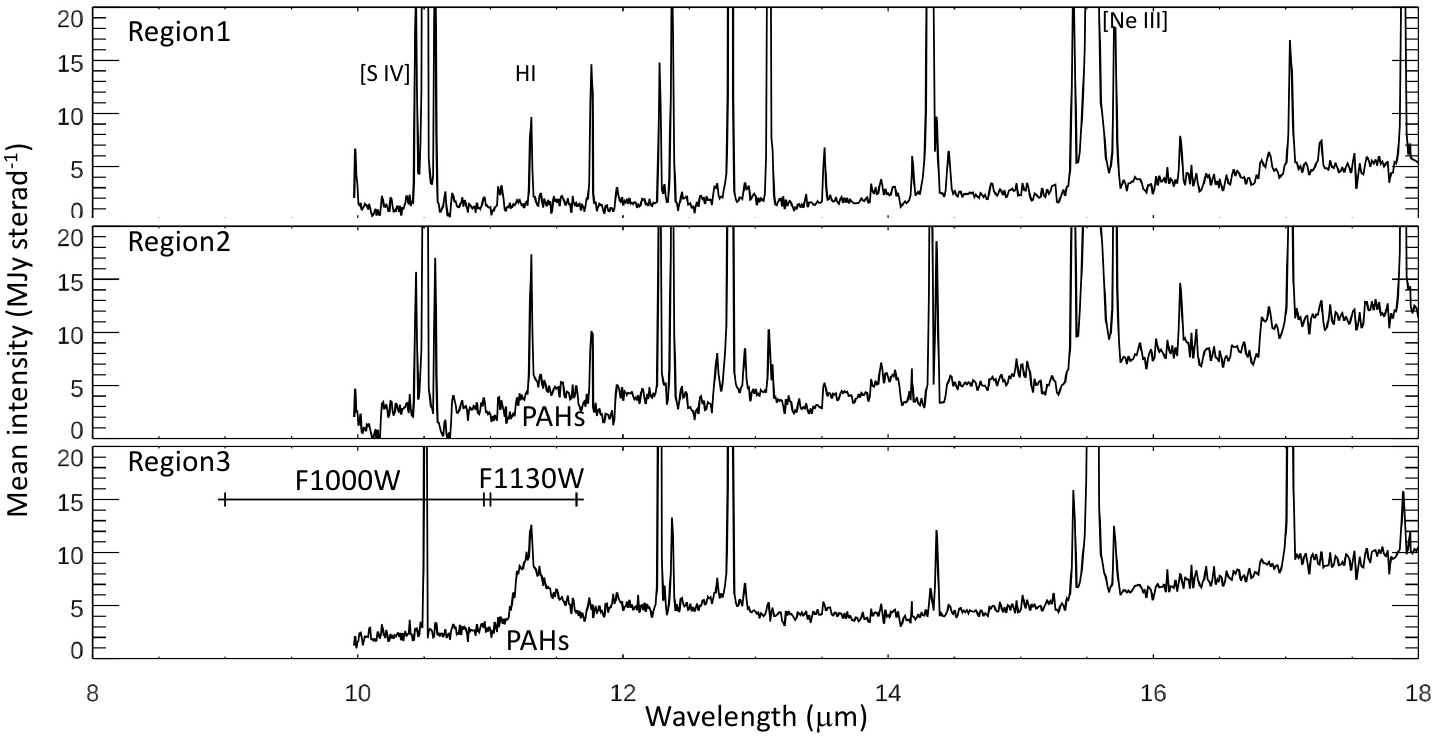}
\caption{ The extracted \textit{Spitzer} spectra  of the Ring Nebula \citep{Cox2016}, from the top towards the inner ring from region 1 to 3. (Fig.~\ref{NIRCam-color-2}).
Region 3 has excess of PAHs at 11.3\,$\mu$m in Figure~\ref{NIRCam-color-2}, and \textit{Spitzer} spectra confirm the presence of 11.3\,$\mu$m PAHs, while region 1, the innermost region of the ring lacks observed PAHs.  
}
  \label{SEDs} 
\end{figure}

Figure~\ref{SEDs} presents the spectra of three different regions across the nebula, from the centre to the west, indicated as regions 1--3 in Figure~\ref{NIRCam-color-2}.
The spectra were extracted from the \textit{Spitzer} 
spectral map specified above.
Region 1, the innermost region of the ring, lacks PAH emission at 11.3\,$\mu$m but shows strong \SIV\, emission in the JWST F1000W filter. Further from the central star, the F1000W image is dominated by continuum emission, as found in the MRS North Region spectra (Table~\ref{observing_log}; van Hoof et al. in preparation).
Regions 2 and 3 both show an excess of the F1130W over the F1000W flux, relative to a simply rising continuum, suggesting the presence of PAH emission. Also for these two fields, the F335M flux has a subtle excess over F300M. These filter bands need cautious interpretations, as both contain several H$_2$ lines. Nevertheless, these excesses appear at the same locations, so we interpret them as being likely due to PAH emission.

The colour-excess ring seen in Figure~\ref{NIRCam-color-2} is centred at the current location of the central star, rather than at the offset centre found for other emission features. The ring also coincides with the thin region at the edge of the shell where the excitation drops off rapidly, as traced by the \NII/\OIII\ ratio \citep{ODell2013}. This suggest that the PAH emission is excited by FUV radiation from the star that penetrates to regions where the nebula has become optically thick to the harder, H-ionizing UV photons, beyond the ionization front.

The PAH emission distribution appears very different from that of the H$_2$ emission which is far more widespread in the nebula. However, we cannot conclude that weaker PAH emission is not present elsewhere in the nebula. Neither can we conclude whether the PAHs are created by the chemistry in the ring or have survived from the molecular AGB wind.


\subsection{Multiplicity}
\label{sec:multi}

A distant companion to the central star was identified by \citet{Gonzalez-Santamaria2021}. This star, \emph{Gaia} DR3 2090486687506009472, lies 18.5~arcsec from the central star, corresponding to a projected distance of 0.07\,pc. It shares both the proper motion and parallax of the central star, and is the only \textit{Gaia} star within 5~arcmin to do so. \citet{Gonzalez-Santamaria2021} proposed that this companion is a white dwarf, based on the \textit{Gaia} DR3 photometry. However, the $B_P$ and $R_P$ photometry is discrepant from the $G$-band photometry by 2.74~mag, and appears to be significantly affected by nebular emission lines. Similar problems appear to affect other optical and near-IR photometry of the star \citep{PanSTARRS,2MASS}, which are discrepant from both the photometry published in the Hubble Source Catalogue \citep{HubbleSourceCatalogue} and from the \textit{Gaia} photometry.

\begin{figure}
    \includegraphics[trim={0.1cm 0.1cm 0.1cm 0.1cm},clip, width=0.50\textwidth]
    {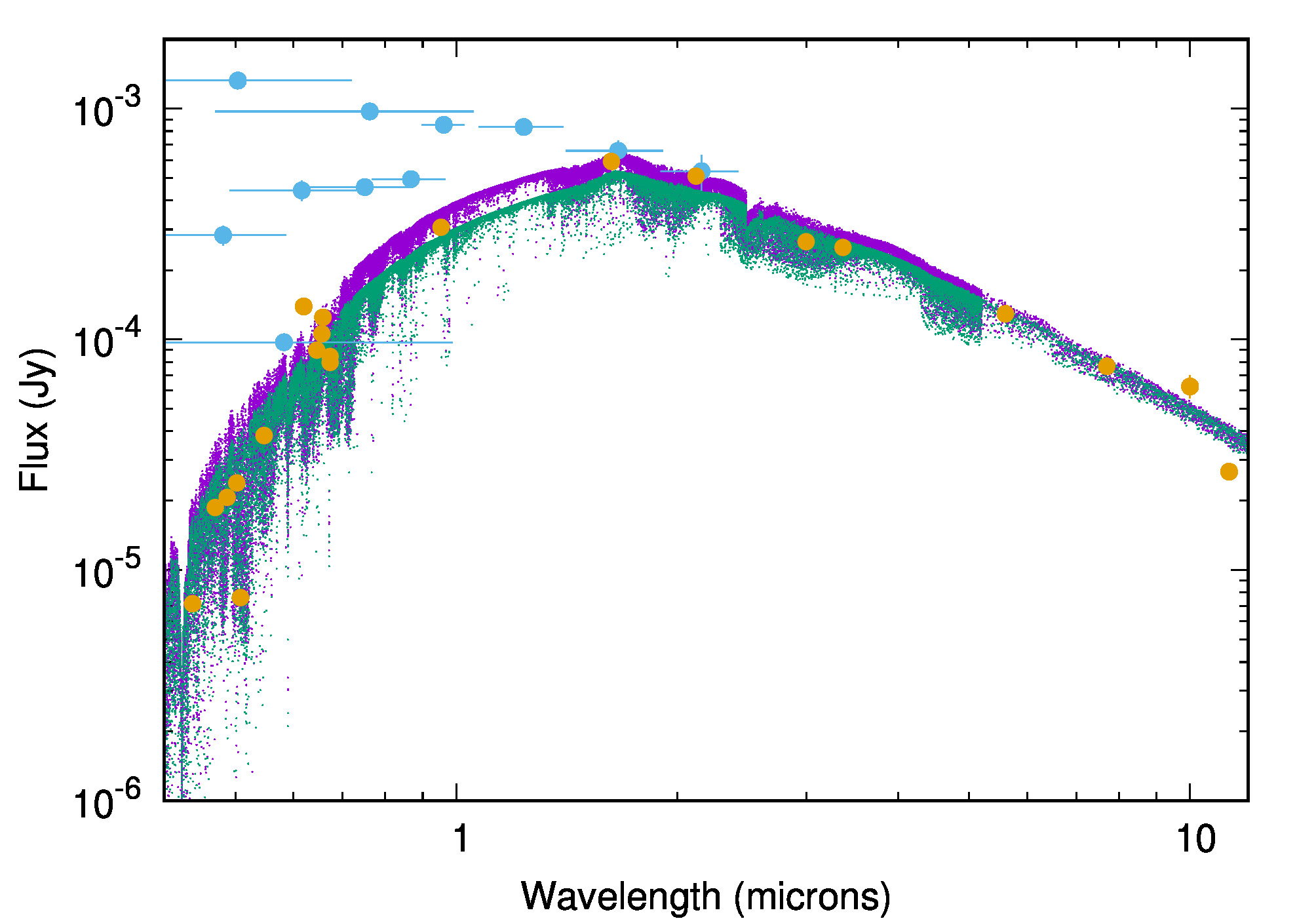}
\caption{ SED of a companion star candidate, \textit{Gaia} DR3 2090486687506009472. Orange points show archival \emph{Hubble Space Telescope} photometry and our new \textit{JWST} photometry. Blue points show other literature data from \textit{Gaia}, the Two Micron All-Sky Survey (2MASS) and the Panoramic Survey Telescope and Rapid Response System (Pan-STARRS) as indicated in the text. Many of these are strongly affected by the bright nebular background. Overlain are {\sc bt-settl} model atmospheres \citep{Allard2014} for 3400 K and 3600 K stars with $\log(g) = 4.5$\,dex, [Fe/H] = --0.5\,dex, [$\alpha$/Fe] = +0.2\,dex and the appropriate reddening.}
  \label{NIRCam-nearbystar} 
\end{figure}

The available photometric data for this star are presented in Fig.~\ref{NIRCam-nearbystar}. This figure includes photometry from our NIRCam and MIRI observations, extracted with an aperture of radius 1~arcsec. The `background' for these \textit{JWST} data points was estimated in an annulus of inner/outer radii of 1.1/1.5~arcsec, with the aperture correction estimated using WebbPSF.

To establish the properties of the companion star, we use the Python Stellar Spectral Energy Distribution toolset (PySSED)\footnote{PySSED is the code underlying the S-Phot software data application. It is currently under development and will be hosted at \url{https://explore-platform.eu/sdas}}, an SED-fitting code based on the software presented in \citet{McDonald2009,McDonald2012,McDonald2017}. We assume a distance to the star of 790~pc, and that the star has a composition of between [Fe/H]\,$ = -0.4$ and +0.0~dex and a corresponding [$\alpha$/Fe] = +0.1 to +0.0~dex, based on its height of 190~pc from the Galactic plane. The Ring Nebula has a metallicity which is approximately solar \citep{Liu2004, Guerrero1997}.

The most significant unknown is the extinction by dust in the nebula at this location, which also depends on whether the star is in front or behind the main shell. We assume a plausible range of $c_{\rm H\beta} = 0.1$ to $0.4$, giving $E(B-V) = c_{\rm H\beta} / 1.46 = 0.068$ to $0.274$~mag. Assuming that the extinction is the dominant source of uncertainty, we fit the parameters listed in Table \ref{tab:companion}.

\begin{table}[]
    \centering
     \caption{{\sc pyssed} parameters of the companion star, for two values of the metallicity}
   \begin{tabular}{cccc}
    \hline
    [Fe/H]          & 0.0  &  $-0.4$  \\ \hline
     $T_{\rm eff}$  & 3281 -- 3406~K  & 3382 -- 3635\,K \\
     $L$  & 0.0136 -- 0.0199~L$_\odot$ & 0.0328 -- 0.0433\,L$_\odot$ \\
     $R$  & 0.362 -- 0.406~R$_\odot$ & 0.525 -- 0.604\,R$_\odot$ \\
     \hline
    \end{tabular}
    \label{tab:companion}
\end{table}

These correspond to a main sequence star of approximate 
spectral type M2--M4 with a mass of 0.3--0.5\,M$_\odot$ \citep{Cifuentes2020}.

At the projected distance of the companion from the central star, the orbital period would be of order $10^6$\,yr and the orbital velocity 250\,m\,s$^{-1}$. If the star was bound before the AGB mass loss, it is likely unbound now. It is plausible that it was originally in a more compact orbit. However, the proper motion shows that it cannot have moved away by more than 3~arcsec during the lifetime of the PN.

Although the companion star lies roughly along the direction of the minor axis, it is not exactly aligned: the offset is about 25\arcdeg. 

The concentric arcs indicate the additional presence of a closer companion, with an estimated period of 280 years, corresponding to a semi-major axis of 50~au for an assumed combined mass of 1.5~M$_\odot$. This is similar to NGC\,3132 where there is also evidence for at least a triple stellar system.

The puzzling 2~arcsec offset between the centre of certain nebular structures and the central star could be related to the multiplicity. 
A speculative model for this involves an elliptical orbit for the close pair. As most of the time in an eccentric orbit is spent near largest separation (apoastron), the nebula is ejected with a typical central velocity corresponding to this part of the orbit. The offset acquired over a dynamical time of 4,000 years corresponds to a velocity of 2\,\kms.  For the system above and assuming an ellipticity of 0.5, 
a velocity difference of the primary component between systemic and aphelion of 2\,\kms~requires an equal mass binary. It should be possible to detect a main sequence star with such a mass, but a white dwarf companion would be difficult to see.

The companion star has remained bound to the central star over its lifetime. Assuming that it formed at this distance, this suggests that no other star has come closer to it than half the current separation of 0.07\,pc. For an assumed number density of stellar systems of 0.1\,pc$^{-3}$, the mean free path is 2600\,pc. Assuming a random velocity of 10\,km\,s$^{-1}$, this gives a time between encounters of 1\,Gyr.  This indicative calculation only favours a higher mass progenitor for the central star of $\gsim 2$\,\Msun\, because of the evolutionary timescales. A more massive system is also more robust against perturbations. Alternatively, the companion may have formed in a tighter orbit and escaped due to the  AGB mass loss of the primary. This requires a mass loss of 50~per~cent\ or more of the original combined mass, which also indicates a primary progenitor mass of 1.5~\Msun\, or more.

\subsection{Comparison to similar planetary nebulae}

The two well-studied PNe most similar to the Ring Nebula  are the so-called Southern Ring Nebula  and the Helix Nebula. 

NGC\,3132 (the Southern Ring Nebula) lies at a similar distance, at 750~pc. It was imaged with \textit{JWST} as part of its Early Release Observations (ERO) programme \citep{pontoppidan2022}. A detailed analysis of these images was carried out by \citet{DeMarco2022}. NGC\,3132 was imaged in fewer filters than NGC\,6720 (the Ring Nebula), but the images show very similar 
structures to those we find here, including the numerous globules visible in H$_2$, the radial spikes in the halo, the system of concentric arcs and the structure inside the cavity.

The Helix Nebula is closer to us, at 200~pc. It too has a system of globules, but the globules in the inner part of its shell have well-defined tails \citep{ODell2004, Matsuura2009}, which are rarely seen in NGC\,6720 and NGC\,3132. Spikes in the halo are detected in the \textit{Spitzer} 5.8~$\mu$m image of the Helix \citep{Hora2006}.

It is interesting to note that the numbers of globules in each of these three nebulae are similar, of order 20\,000 \citep{Meixner2005, DeMarco2022}. Applying the same peak-finding algorithm that we used to count the clumps in the Ring Nebula (Section~\ref{clumpfindingtext}) to the Southern Ring Nebula we find 15\,000 clumps.  However, there are notable differences in the size and shape of globules in the Ring Nebula compared to the other two objects. The Helix globules typically have a diameter of 2~arcsec in H$_2$ images \citep{Matsuura2009} at a distance of 200~pc, and one of the largest globules in the Southern Ring Nebula is 0.5~arcsec in diameter at a distance of 750~pc \citep{DeMarco2022}. These angular diameters correspond to $\sim$400~au. In contrast, the  globules in the Ring Nebula tend to be smaller. One of the largest globules  in the Ring Nebula is indicated in Fig.~\ref{components}. The diameter of the head is approximately 0.4~arcsec across in the F212N H$_2$ image, corresponding to 300~au.  The typical diameter of the globules is 0.2 arcsec, or 150 au. Most of the Ring Nebula globules also lack significant tails.

Another similar PN of relevance is NGC\,2346, which has been imaged in H$_2$ from the ground \citep{Manchado2015}. It has a more flaring bipolar structure with a thinner torus than the Ring Nebula, but the torus shows similar H$_2$ clumps. The clumps range in size from  0.16 to 0.34~arcsec which, at a \textit{Gaia} DR3 distance of 1400\,pc, corresponds to 200 to 500\,au. This is larger than for the Ring Nebula but similar to those in the Southern Ring and the Helix Nebula.

The formation mechanism of the globules is still under debate. 
Globules may form before the ejection of the PN, and survive in the medium \citep{Zanstra1955}, 
or alternatively, they may have formed during the PN phase \citep{Capriotti1973, Garcia-Segura2006}.
The central stars of these four PNe are all on the cooling track of the HR diagram. This supports a model in which the globules form after the rapid fading of the central star on the cooling track of the white dwarf phase, in the recombining dense gas of the shell \citep{ODell2007,vanHoof2010}.
Further support for a picture where clump formation occurs predominantly at late times, rather than before the ejected envelope has been ionized, was found by \citet{Huggins2002}, who demonstrated the absence of small-scale structures in objects in the early stages of transition from proto-planetary nebulae to full-fledged PNe.

During recombination, the recombination rate ($t_{\rm rec}$) is faster in denser regions ($t_{\rm rec}= 10^5\,{\rm yr}/n_e $ for hydrogen). 
If density fluctuations are present, the higher-density regions will recombine first, lose electron pressure, and collapse under the pressure of the surrounding ionized gas, which will be $\sim$200 times higher than in the recombined regions. Molecules and dust can form, shielded and shielding against the UV radiation. The effect of this shielding can be seen via extinction by the neutral globules against the ionized gas background emission and shadowing in the tails.

The tails in the globules of the Helix nebula are best explained by shadowing \citep{Canto1998, Andriantsaralaza2020} within the photoionized region, triggering recombination and allowing CO to form  along the tail \citep{Andriantsaralaza2020}.  Involvement of stellar wind ablation in tail formation \citep{Dyson2006, 2007MNRAS.382.1447M} cannot be excluded, but stars on the white dwarf cooling track do not have strong stellar winds. 

The lack of tails in the Ring nebula may point to it being in an earlier stage of evolution. The shell of the Helix Nebula is larger than those of the other nebulae, at a radius of 0.33~pc \citep{ODell2004} versus 0.1~pc for NGC 6720 and 0.07~pc for NGC 3132.  The Helix Nebula has a larger dynamical age of $\sim$\,7,400 years \citep{Gonzalez-Santamaria2021} and has likely been on the cooling track for longer. 

On the cooling track, the star initially fades rapidly, leading to recombination. But the fading slows down dramatically after that \citep{MillerBertolami2016}. During this phase, the nebula keeps expanding and the density drops. Once the star stops fading, the ionization front should move outward through the nebula again, due to the expansion of the nebula and hence lower density and longer recombination time scales. The ionization front  can now pass the previously formed globules. Once that happens, the globules may develop tails in the shadowed region in the re-ionized gas. This could explain why the inner globules in the older Helix Nebula have tails, while those in the younger nebulae do not.

\subsubsection{Hydrodynamical model}

\begin{figure}
    \includegraphics[trim={0.2cm 0.2cm 0.2cm 0.2cm},clip, width=0.49\textwidth]{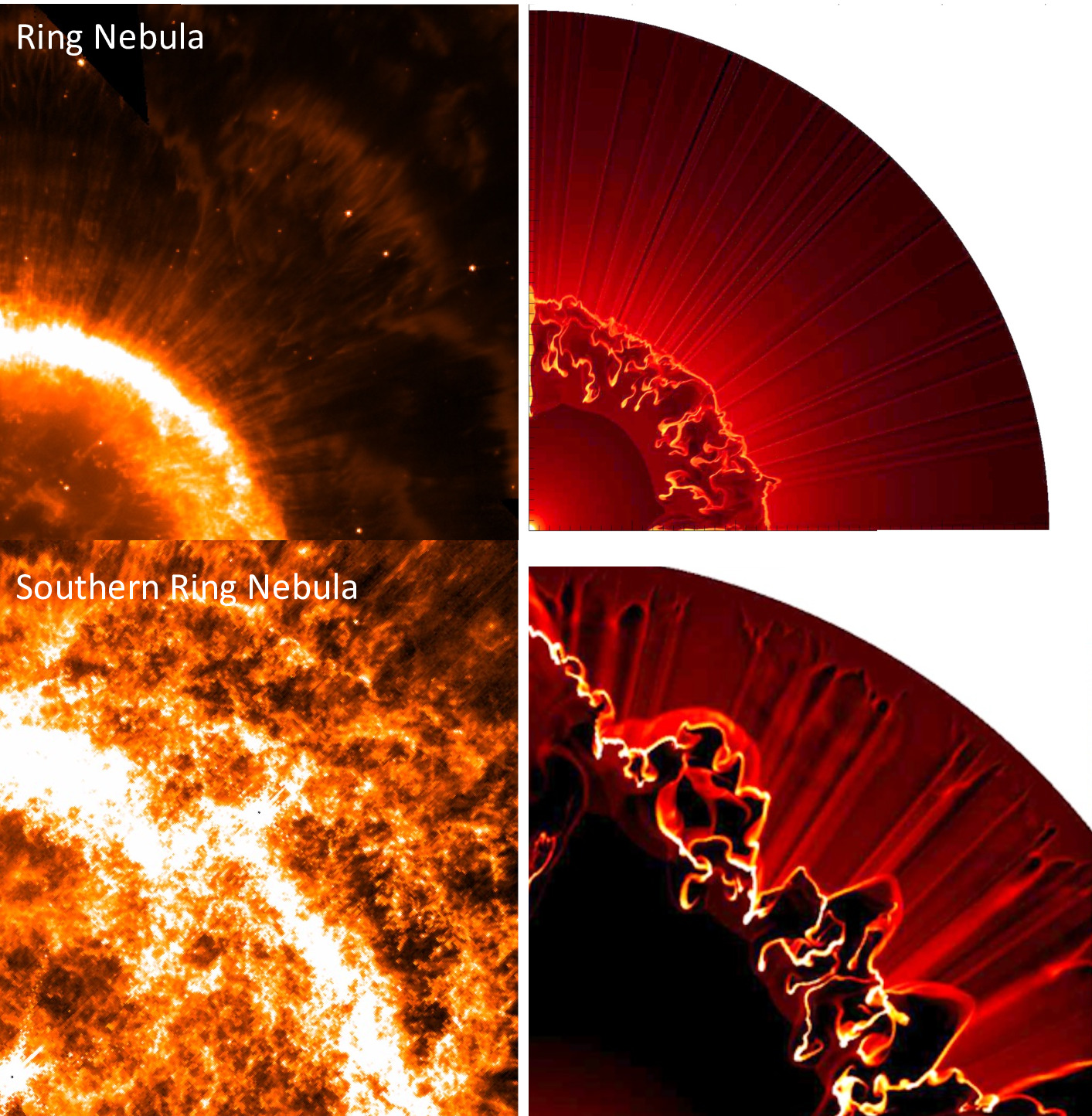}
\caption{ (Right) Density snapshot of a hydrodynamic  simulation for the Ring Nebula (right top) and the Southern Ring Nebula (right bottom). The hydrodynamic model for the Southern Ring Nebula is taken from \citet{DeMarco2022}. 
(Left) \textit{JWST} images of the Ring Nebula (top left; F770W) and the Southern Ring Nebula (bottom left; F212N), highlighting the spikes and clumps. The
Southern Ring Nebula has evolved faster, and has more fully developed spikes and clumps.
  \label{hydro} }
\end{figure}

Figure~\ref{hydro} shows density snapshots from hydrodynamical models \citep{Garcia-Segura2006, Garcia-Segura2018} for the Ring Nebula and the Southern Ring Nebula. The models are extracted for an age of 4000\,yr after envelope ejection, a time equal to the dynamical age of the Ring Nebula \citep{ODell2013}. The 2-d simulation has a resolution of 1000$times$1000 zones in spherical polar
coordinates (R, $\Phi$) on the orbital or equatorial plane. The model assumes ejection via a common envelope evolution event which provides an equatorial density enhancement. There is currently no evidence that the Ring Nebula underwent a common envelope event, since the posited binary companions are too distant to lead to a common envelope phase. However, the  common envelope leads to formation of a high-density torus which is the structure of the shell of the Ring Nebula. The model follows the expansion of the ejected shell, and includes central star evolution \citep{Vassiliadis1994} and ionization of the nebula.

The model (Fig.~\ref{hydro}) shows the formation of higher density regions in the form of clumps in a way similar to described above. They are formed due to the thin-shell instability in the swept-up shell. The thin-shell instability acts in wind-blown bubbles \citep{Vishniac1983} and is very effective at producing clumps \citep{Garcia-Segura1995}. It may involve photo-ionization. The number of clumps depends on the thickness of the swept-up shell, which is related to the temperature and pressure of the shell, and also depends on the expansion velocity. 

Another instability can occur in thicker shells sandwiched between an ionization front and a shock front, which can fragment through a so-called I-S (Ionization-shock) instability \citep{GarciaSegura1996}. This does not involve a wind. in the I-S instability the clumps form at the ionization front. In the model, the first mechanism acts, but both mechanisms can lead to clumping.  

 The clumps are in pressure equilibrium. This may provide a natural explanation for the clumps being larger in the Helix Nebula: it has expanded further with lower density in the ionized region, allowing the clumps to expand.

The I-S instability can form tails directly behind the clumps, caused by direct shadowing. The thin-shell mechanism does not form tails connected to the clumps, because of the  hot-shocked gas and internal gas motions in the bubble. However, once the wind stops, tails can form progressively while the hot-shocked gas disappears, in the same way as for the I-S instability. 

 Fig.~\ref{hydro} compares the Ring Nebula with the Southern Ring nebula, with models for each \citep{DeMarco2022}. The model for the Southern Ring corresponds to a slightly later phase in the evolution.

\subsection{Spikes}

Narrow spikes are found in the halo of the Ring Nebula, the Southern Ring, and the Helix. Although rarely reported  \citep[e.g. NGC\,6543;][]{Guerrero2020}, they may be a common phenomenon in PNe. They are seen in H$_2$ emission in the Ring Nebula,  the Southern Ring \citep{DeMarco2022} and the Helix Nebula \citep{Hora2006} but in optical emission in NGC~6543 \citep{Guerrero2020}. The latter paper presents a hydrodynamical model where the spikes appear to form behind holes in the dense shell.

The fact that they are well aligned with the star shows the importance of illumination. There does not appear to be a direct relation between the spikes and the clumps; the spikes number only around 2~per~cent the number of clumps, and the clumps in the Ring Nebula do not have tails.

The spikes in the Ring Nebula are seen in H$_2$. They are likely excited by radiation passing through the shell.
While the clumpy gas in the shell is optically thick for $<$912~\AA\  radiation which can photoionize hydrogen, it can still be optically thin at $>$912~\AA. Thus, longer wavelength UV radiation can escape through the swept-up shell, exciting H$_2$ molecules via fluorescence in the remnant AGB wind.  

The H$_2$ molecules in the halo are unlikely to result from recombination of the ionized gas, since recombination times in the halo are much longer than those in the main shell. The spikes therefore likely formed already during the high-luminosity PN phase on the horizontal track, and were shadowed during this phase which allowed the original molecules to survive while the surrounding halo became ionized. 

The hydrodynamical model described above also produces spikes. They form in the halo where there is no hot shocked gas, again by shadowing. The lack of hot gas explains why spikes formed but tails did not. There are open issues: the model predicts a similar number of spikes and clumps, and because of the high density it has fast recombination.

Comparison of this work with that for the Southern Ring nebula \citep{DeMarco2022}  shows that the spikes are less developed in the Ring Nebula.  The mass of the progenitor star may play a role, as the central star of the Ring Nebula was likely of lower initial mass ($\sim 1.5$--2 \Msun) than that of the Southern Ring ($2.86 \pm 0.06\,\rm$ \Msun). 

The total numbers of globules in these three PNe are similar, of order 20\,000. 
This may be because the densities and expansion velocities of the nebulae were similar, as these two parameters are key to triggering the instabilities that initiated the formation of the globules.

\subsection{Halo}

The \textit{JWST} images  show the presence of H$_2$ in the halo, for example in the F335M filter. The images show a combination of azimuthal structures further out and flocculent structure in the inner halo. The latter is mainly seen closer to the major axis, outside the shell.

\begin{figure}
    \centering
    \includegraphics[width=8.5cm]{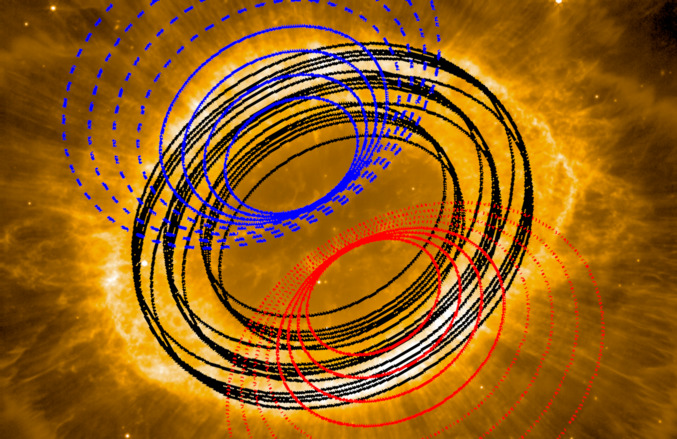}
    \includegraphics[width=8.5cm]{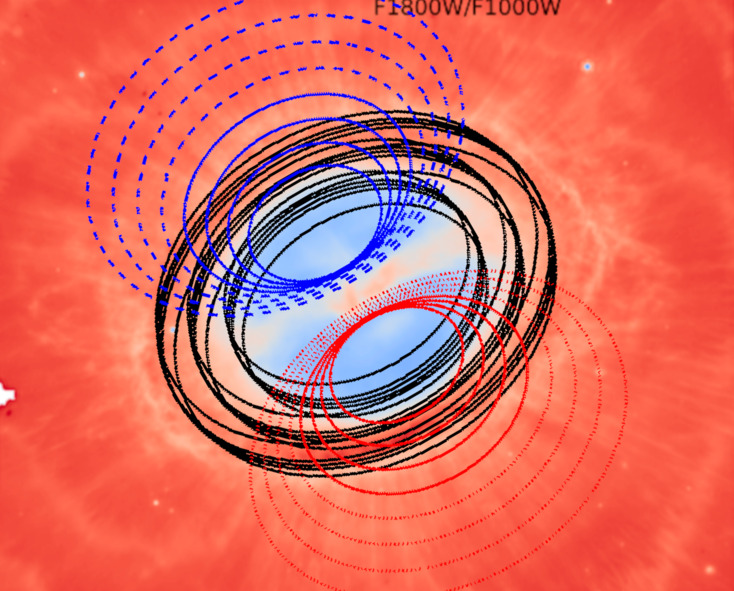}
    \caption{Top: The F770W images with the contours of the bicone superposed. Black lines show the torus, blue and red the cones. The dashed lines show the cone in the halo. Bottom: the same model superposed on the F1800W (red) / F1000W (blue) ratio image}
    \label{fig:cone}
\end{figure}

\citet{Bryce1994} proposed a biconical structure, with two lobes extending into the halo (their Fig. 11). This model was refined by \citet{Sahai2012} and \citet{ODell2013}. The bicone has a wide opening angle, and is confined by the torus which forms the shell. The polar axis is close to the line of sight, making the bicone not obvious in the images. The projected bicone is seen in long-slit spectra, however, due to its higher flow velocity.

We created a morphological model consisting of a barrel-like torus, and a wide bipolar flow angled 30 degrees from the line of sight, towards the minor axis of the shell. The lobes start at the shell and move out into the inner halo. The lobe is assumed to be slightly flaring, with radius $r \propto z^{1.5}$. The  H$_2$ is assumed to be located  on the wall of the cone, swept out from the shell. This model is speculative.

Fig. \ref{fig:cone} shows the F770W \textit{JWST} image, with the contours of the wall of the cone superposed. In this model, the two cones are slightly separated in projection. Where the contours run close together, there is a longer line of sight along the cone and more H$_2$ emission may be expected.  The contours roughly agree with the location of the H$_2$. Interestingly, it can also explain the two stripes seen in projection against the cavity which become the edge of the bicone. Finally, the arcs and stripes are both seen where the wall of the cones is seen under the most favourable angle, around the polar axis. 

The model is overplotted on the F1800W/F1000W ratio image in the bottom panel of Fig. \ref{fig:cone}. F1800W is dominated by \SIII\ and F1000W by \SIV. The latter is seen in the openings of the cones which provide a clear line of sight into the cavity. The lower excitation \SIII\ comes from the shell. The region between the stripes also shows this emission from the shell, projected on the cavity, and therefore has a very different appearance in the ratio image.

\section{Conclusion}

The \textit{JWST} images have revealed a wealth of structural detail in NGC 6720. The nebula has a highly ionized inner cavity, a shell in which some 20\,000 dense clumps contain up to half the total mass, a thin ring of possible PAH emission, and a halo that contains around 10 concentric arcs and 400 spikes. The centre of the nebula is offset by 2~arcsec from the central star. Much of this detail is shown by the H$_2$ emission. 

The globules/clumps have densities of $n_{\rm H} \sim 10^5$--$10^6$ cm$^{-3}$ and account for $\sim 0.1$\,\Msun, up to half the mass of the PN. They are modelled as arising from thin-shell or I-S instabilities The PNe in which clumps are seen have  central stars that are already on the cooling track, suggesting that the clumps form during the rapid fading of the star. The globules in the Ring Nebula have little or no tails, unlike those in the Helix Nebula. The globules in the Helix nebula are also larger. These differences could be due to the Helix Nebula being more evolved and the Ring nebula being in an earlier phase of evolution. 

The radial spikes in the halo are seen in H$_2$. They are regions partially shadowed by the shell, predicted by the hydrodynamical model.

The central exciting star is inferred to be a member of a triple system. This consists of: the central star itself, with a progenitor mass of $\sim$1.5--2\,\Msun; a binary companion at some 35\,au, responsible for the evenly spaced concentric arc structures in the nebula; and a distant, common proper motion companion at 0.07\,pc which is inferred to be a low-mass M2--M4 main sequence star.

A schematic model is presented which consists of the central torus and two polar cones, in which the H$_2$ is swept up on the walls of the cones. The model can explain the location of the H$_2$ emission in the halo and the stripes seen on projection against the cavity. 
 
 Many features we see in the \textit{JWST} images of the Ring Nebula, including the spikes, are shared by several other well-studied PNe of similar morphology. The time when planetary nebulae could be modelled as uniform density spheres is long gone. They contain a large variety of structures and phases, from highly ionized hot gas to dense molecular clumps.

\section*{Acknowledgements}
This work is based on observations made with the NASA/ESA/CSA James Webb Space Telescope. The data were obtained from the Mikulski Archive for Space Telescopes at the Space Telescope Science Institute, which is operated by the Association of Universities for Research in Astronomy, Inc., under NASA contract NAS 5-03127 for \textit{JWST}. These observations are associated with program \#1558.

Based on observations made with the NASA/ESA Hubble Space Telescope, and obtained from the Hubble Legacy Archive, which is a collaboration between the Space Telescope Science Institute (STScI/NASA), the Space Telescope European Coordinating Facility (ST-ECF/ESAC/ESA) and the Canadian Astronomy Data Centre (CADC/NRC/CSA).

This work has made use of data from the European Space Agency (ESA) mission {\it Gaia} (\url{https://www.cosmos.esa.int/gaia}), processed by the {\it Gaia} Data Processing and Analysis Consortium (DPAC, \url{https://www.cosmos.esa.int/web/gaia/dpac/consortium}). Funding for the DPAC has been provided by national institutions, in particular the institutions participating in the {\it Gaia} Multilateral Agreement.

This work is based in part on observations made with the \textit{Spitzer Space Telescope}, which was operated by the Jet Propulsion Laboratory, California Institute of Technology under a contract with NASA.

This study is based on the international consortium of ESSENcE (Evolved Stars and their Nebulae in the JWST era).

R.W. and M.M. acknowledge support from STFC Consolidated grant (2422911).
M.J.B. and R.W. acknowledge support  from European Research Council (ERC) Advanced Grant SNDUST 694520. 

This research has used data, tools or materials developed as part of the EXPLORE project that has received funding from the European Union’s Horizon 2020 research and innovation programme under grant agreement No 101004214. A.A.Z.\, I.M.\, and N.L.J.C.\, acknowledge support from this grant.

A.A.Z. acknowledges funding through UKRI/STFC through grant ST/T000414/1.

I.A. acknowledges support from the Coordena\c{c}\~{a}o de Aperfei\c{c}oamento de Pessoal de N\'{i}vel Superior - Brasil (CAPES; Finance Code 001) and the Program of Academic Research Projects Management, PRPI-USP.

H.L.D. acknowledges support from grants JWST-GO-01558.03 and NSF AAG-1715332.
 
G. G.-S. thanks Michael L. Norman and the Laboratory for Computational Astrophysics for the use of ZEUS-3D. The computations were performed at the Instituto de Astronom{\'i}a-UNAM at Ensenada.

P.J.K. acknowledges support from the Science Foundation Ireland/Irish Research Council Pathway programme under Grant Number 21/PATH-S/9360.

RS’s contribution to the research described here was carried out at the Jet Propulsion Laboratory, California Institute of Technology, under a contract with NASA.

J.C., N.C. and E.P. acknowledge support from the University of Western Ontario, the Institute for Earth and Space Exploration, the Canadian Space Agency (CSA)[22JWGO1-22], and the Natural Sciences and Engineering Research Council of Canada.

This research made use of {\sc photutils}, an {\sc astropy} package for
detection and photometry of astronomical sources (\citealt{larry_bradley_2023_7946442}).

\section*{Data Availability}

\textit{JWST} data are available from the Barbara A. Mikulski Archive for Space Telescopes (MAST; \url{https://mast.stsci.edu}). Reduced images will be available by 
request to the authors.



\bibliographystyle{mnras}
\bibliography{ring_jwst} 




\appendix

\section{NIRCam images}

\begin{figure*}
  \begin{minipage}[c]{1\textwidth}
    \includegraphics[trim={0.2cm 0.2cm 0.2cm 0.2cm},clip, width=0.5\textwidth]{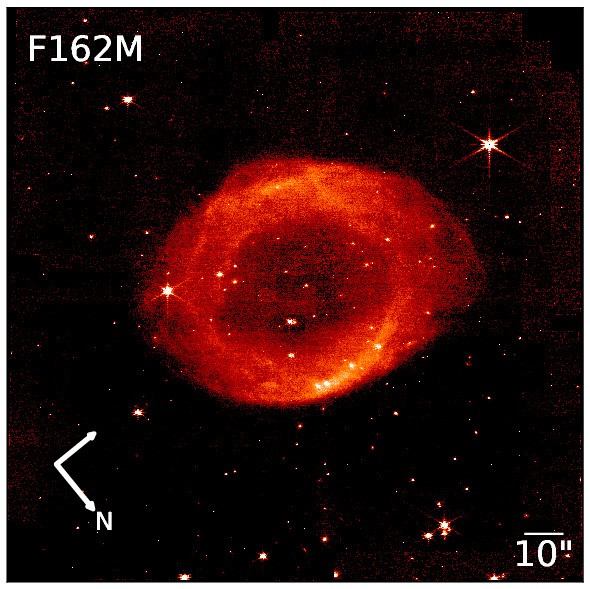}
    \includegraphics[trim={0.2cm 0.2cm 0.2cm 0.2cm},clip, width=0.5\textwidth]{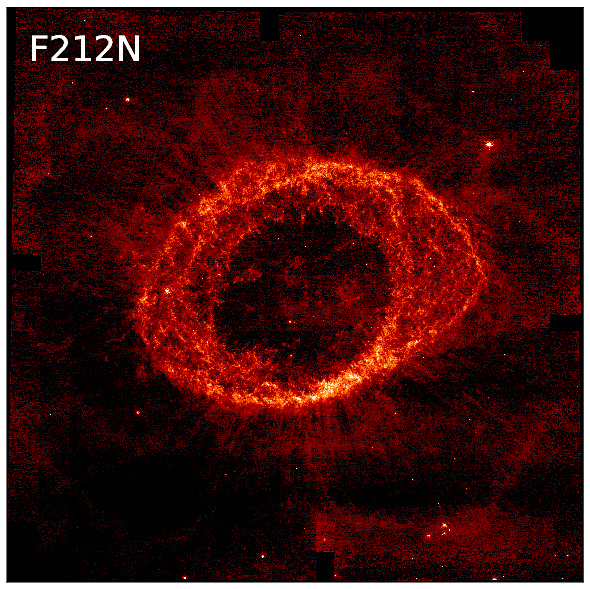}
    \includegraphics[trim={0.2cm 0.2cm 0.2cm 0.2cm},clip, width=0.5\textwidth]{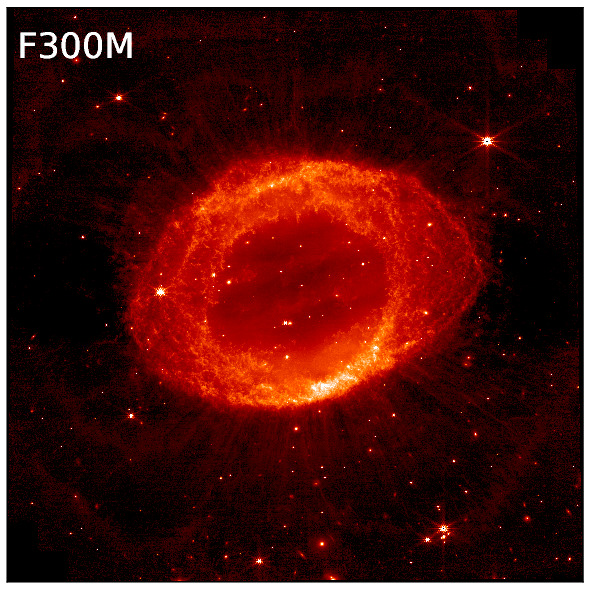}
    \includegraphics[trim={0.2cm 0.2cm 0.2cm 0.2cm},clip, width=0.5\textwidth]{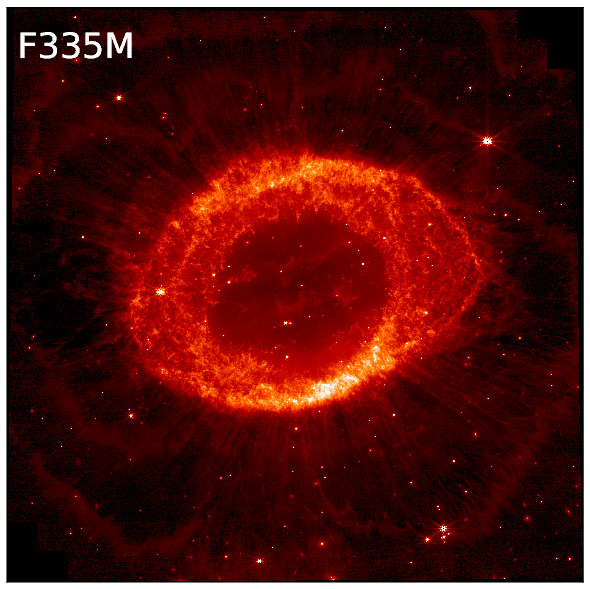}
      \end{minipage}\hfill
\caption{ NIRCam individual images
}
  \label{NIRCam-individual} 
\end{figure*}

Figure~\ref{NIRCam-individual}  shows NIRCam  images with individual filter bands. The F162M image is distinctly different from the rest of NIRCam images, being bright in the ring, with little emission from halo. This filter
contains continuum, presumably, free-free emission, and hydrogen recombination lines, and these are fainter than molecular hydrogen emission , which dominates the other three NIRCam images.

\section{Flux ratio images}

\begin{figure}
  \begin{minipage}[c]{1.0\textwidth}
    \includegraphics[trim={6.5cm 2.5cm 3.8cm 2.5cm},clip, width=0.50\textwidth]{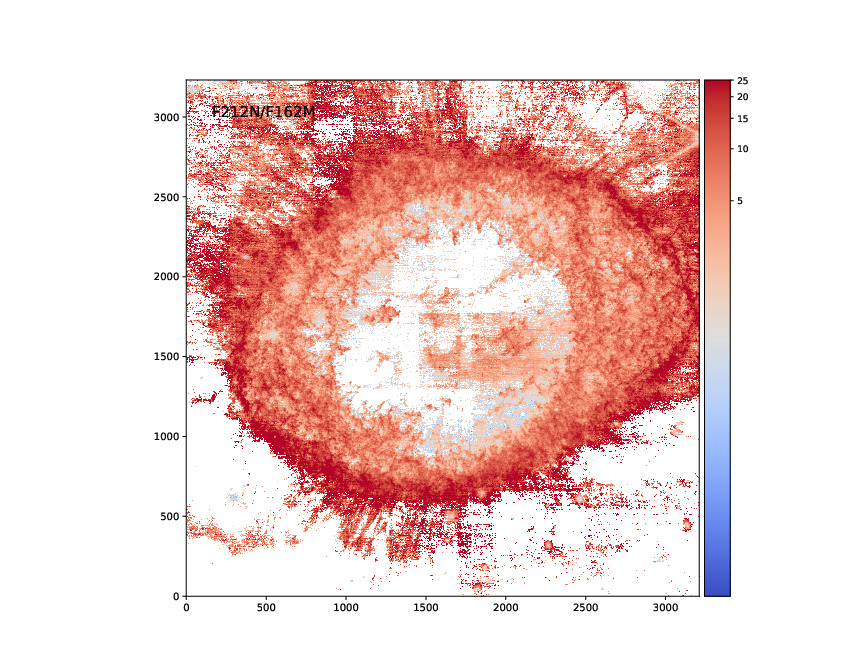}
       \end{minipage}\hfill
\caption{ F212N/F162M ratio in log scale.
}
  \label{NIRCam-color-1} 
\end{figure}

\begin{figure*}
  \begin{minipage}[c]{1\textwidth}
    \includegraphics[trim={2.6cm 3.5cm 0.8cm 4.2cm},clip, width=0.50\textwidth]{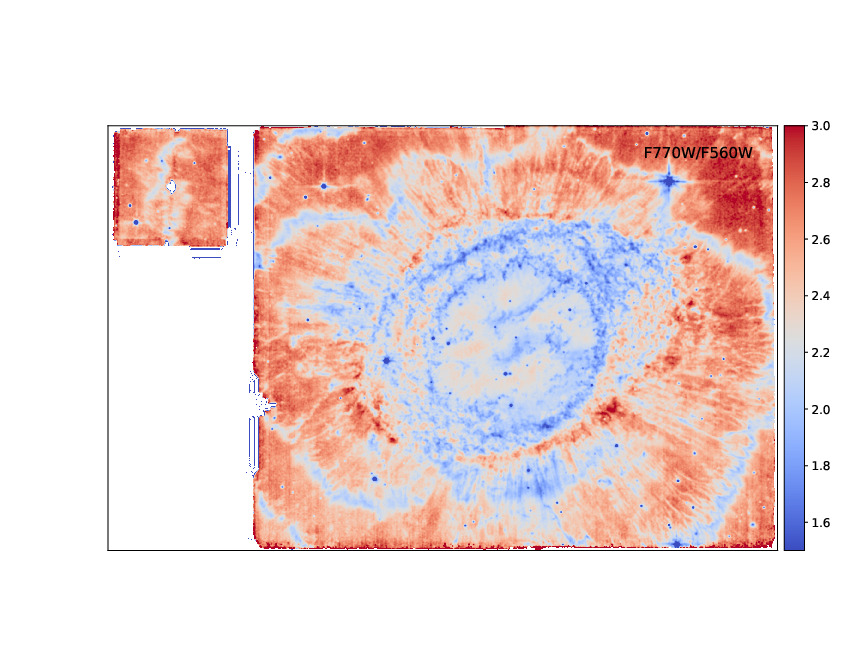}
    \includegraphics[trim={2.6cm 3.5cm 0.8cm 4.2cm},clip, width=0.50\textwidth]{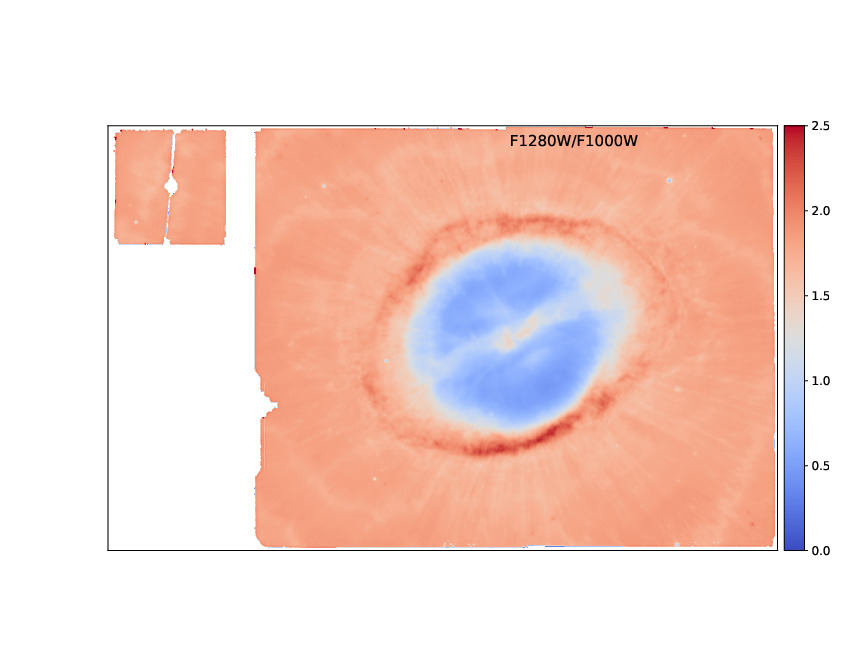}
    \includegraphics[trim={2.6cm 3.5cm 0.8cm 4.2cm},clip, width=0.50\textwidth]{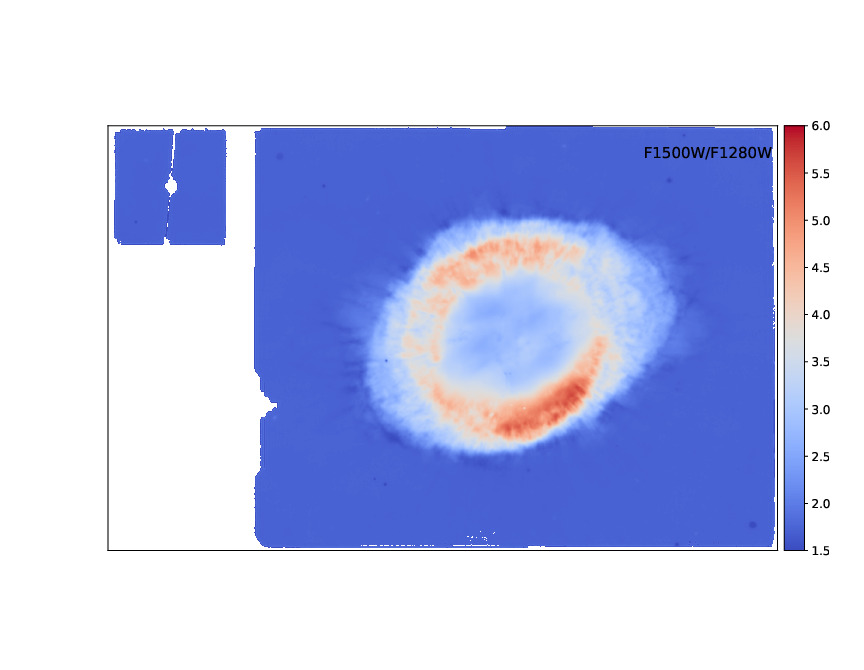}
    \includegraphics[trim={2.6cm 3.5cm 0.8cm 4.2cm},clip, width=0.50\textwidth]{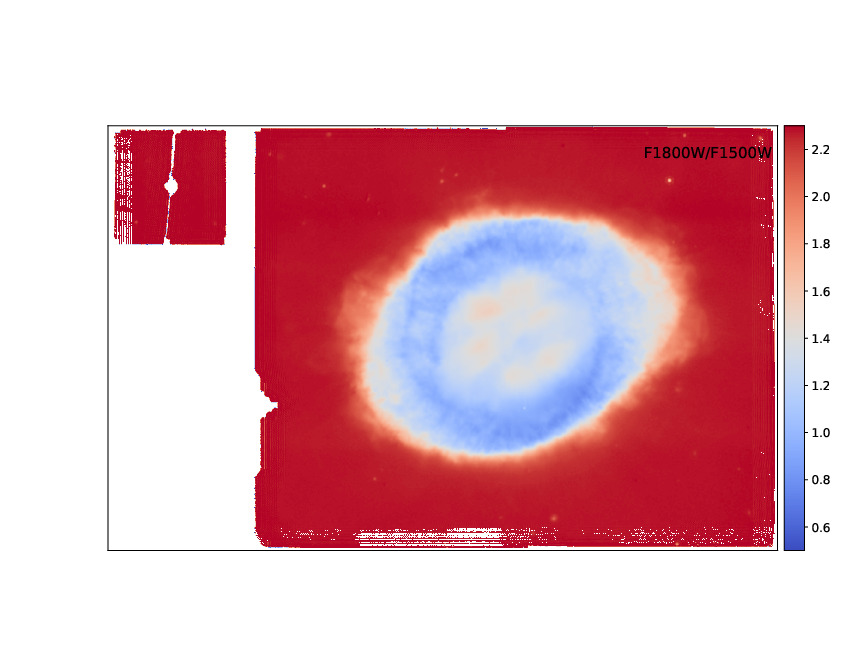}
    \includegraphics[trim={2.6cm 3.5cm 0.8cm 4.2cm},clip, width=0.50\textwidth]{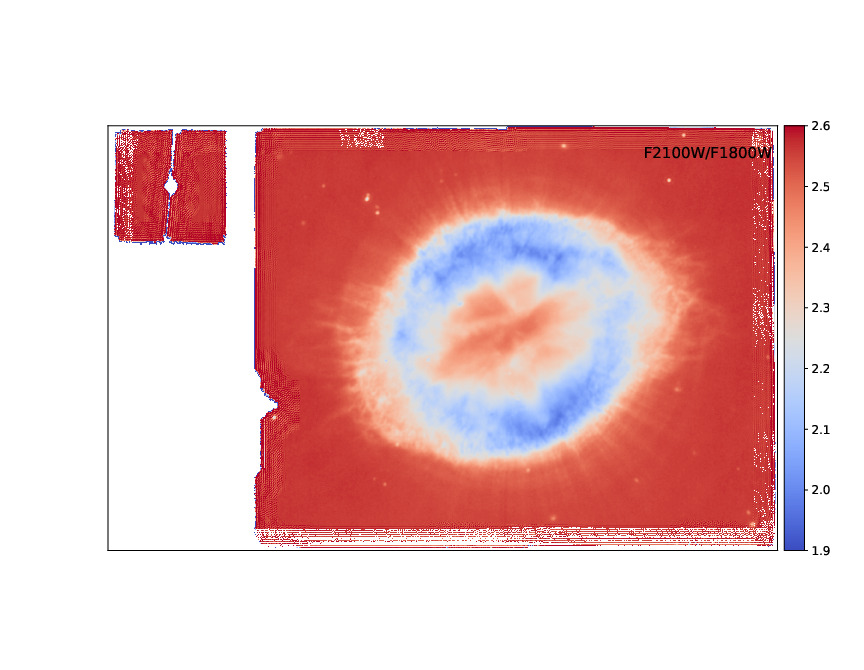}
    \includegraphics[trim={2.6cm 3.5cm 0.8cm 4.2cm},clip, width=0.50\textwidth]{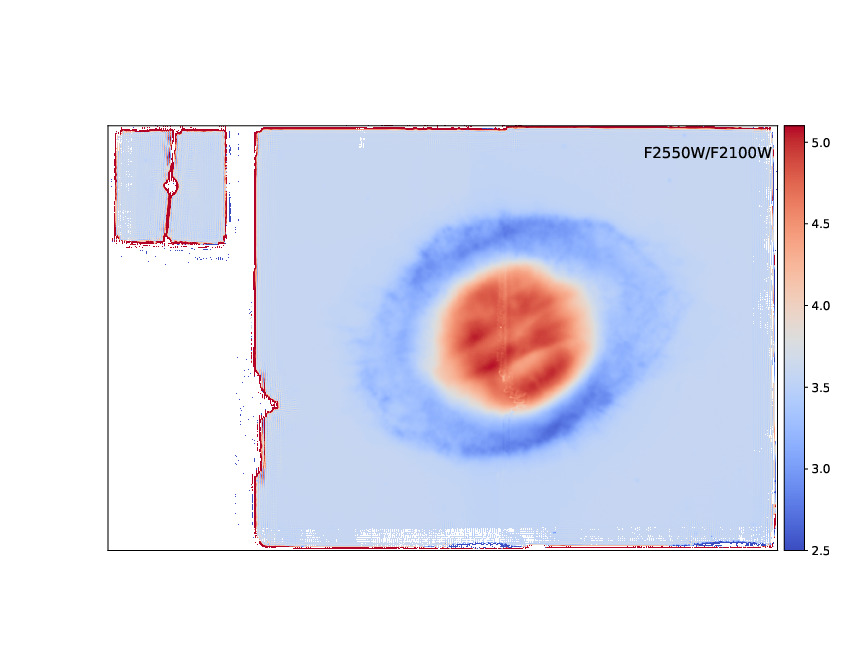}
       \end{minipage}\hfill
\caption{ MIRI colour (flux ratio) images.
}
  \label{MIRI-color} 
\end{figure*}

\begin{figure}
  \begin{minipage}[c]{1\textwidth}
    \includegraphics[trim={8.2cm 2.cm 6.0cm 2.5cm},clip, width=0.50\textwidth]{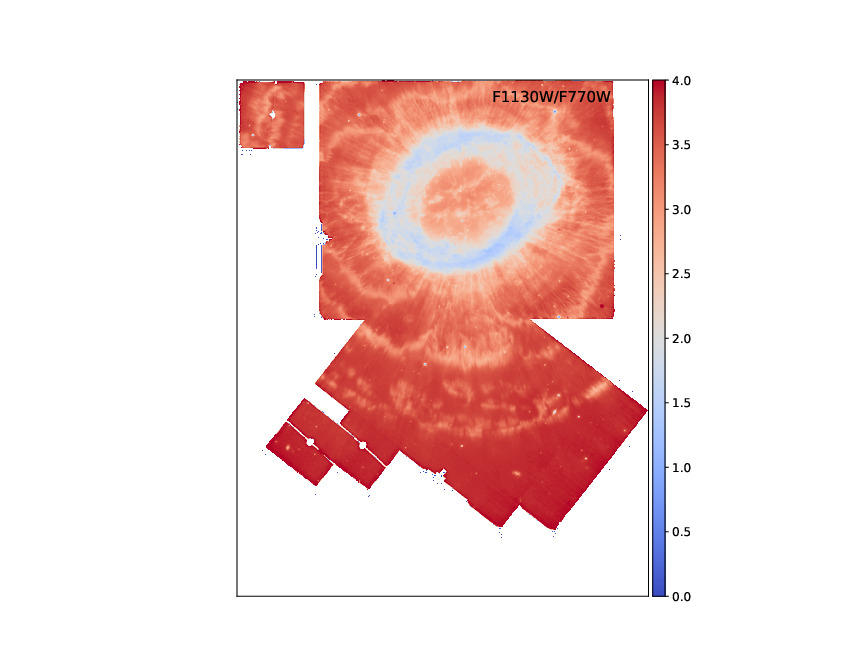}
       \end{minipage}\hfill
\caption{MIRI colour of F1130W/F770W, which includes the extension towards the north-west.
}
  \label{MIRI-color-extended} 
\end{figure}

Figures~\ref{NIRCam-color-1}--\ref{MIRI-color-extended}   show flux ratio images of pairs of filter bands, demonstrating the contributions of different lines/continuum or dust at different locations.

Figure~\ref{NIRCam-color-1} demonstrates that H$_2$ emission is found in many globules, and across the halo. Due to its large dynamical range, this image only is plotted in log scale

Figure~\ref{NIRCam-color-2} shows the F335M/F300M image. Similar to Figure~\ref{NIRCam-color-1}, the F335M/F300M colour is reddest at the outer edge of the shell.
These two filter bands contain more or less similar components (Table~\ref{observing_log}). This red edge at the shell may be explained by the additional contribution of PAHs to the emission in the F335M filter passband. If it is due to different H$_2$ line ratio, or continuum, the colour should be redder as further away from the central star, but that is not found in this figure. Hence, most likely PAHs are contributing to F335M at the edge of the shell, where UV radiation from the central star may fall onto the appropriate conditions for UV excitation.

Figures~\ref{MIRI-color} and \ref{MIRI-color-extended} show MIRI flux ratio images.
MIRI F560W--G1130W images tend to have substantial contributions from the continuum, with additional line contributions on top.
The F770W/F560W and F1130W/F770W images show similar colour patterns: the shell has rather similar colour within, in contrast to the F335M/F300M image.
The F770W and F560W filters contain similar emission-line contributions, with additional \ion{Ar}{ii} emission in F770W.
This additional \ion{Ar}{ii} contribution probably characterises the blue stripe structure inside the cavity, and the cavity itself.
The edge of the shell is rather red in F770W/F560W, which may suggest a contribution from PAHs at this location.
F1130W/F1000W looks very similar to F335M/F300M with the presence of red colour at the edge of the shell. 
Considering that F1130W has the least line contribution amongst all filter bands, having the reddest colour at the shell edge could be due to PAHs, while the red colour in the halo might be due to the contribution from cold dust emission \citep{Cox2016}.


\section{Filter transmission curves}

\begin{figure*}
  \begin{minipage}[c]{1\textwidth}
    \includegraphics[trim={3.0cm 2.cm 3.3cm 2.2cm},clip, width=0.50\textwidth]{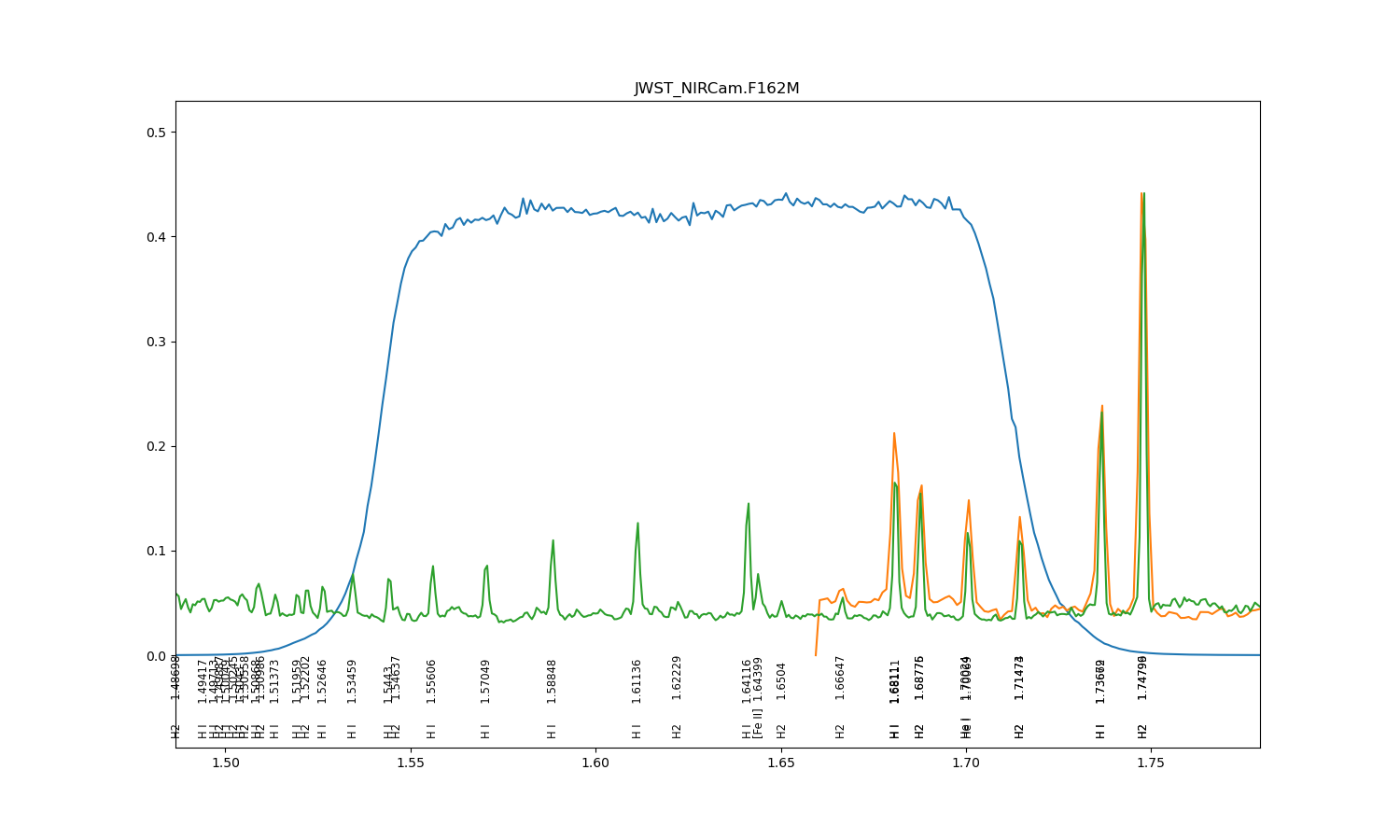}
    \includegraphics[trim={3.0cm 2.cm 3.3cm 2.2cm},clip, width=0.50\textwidth]{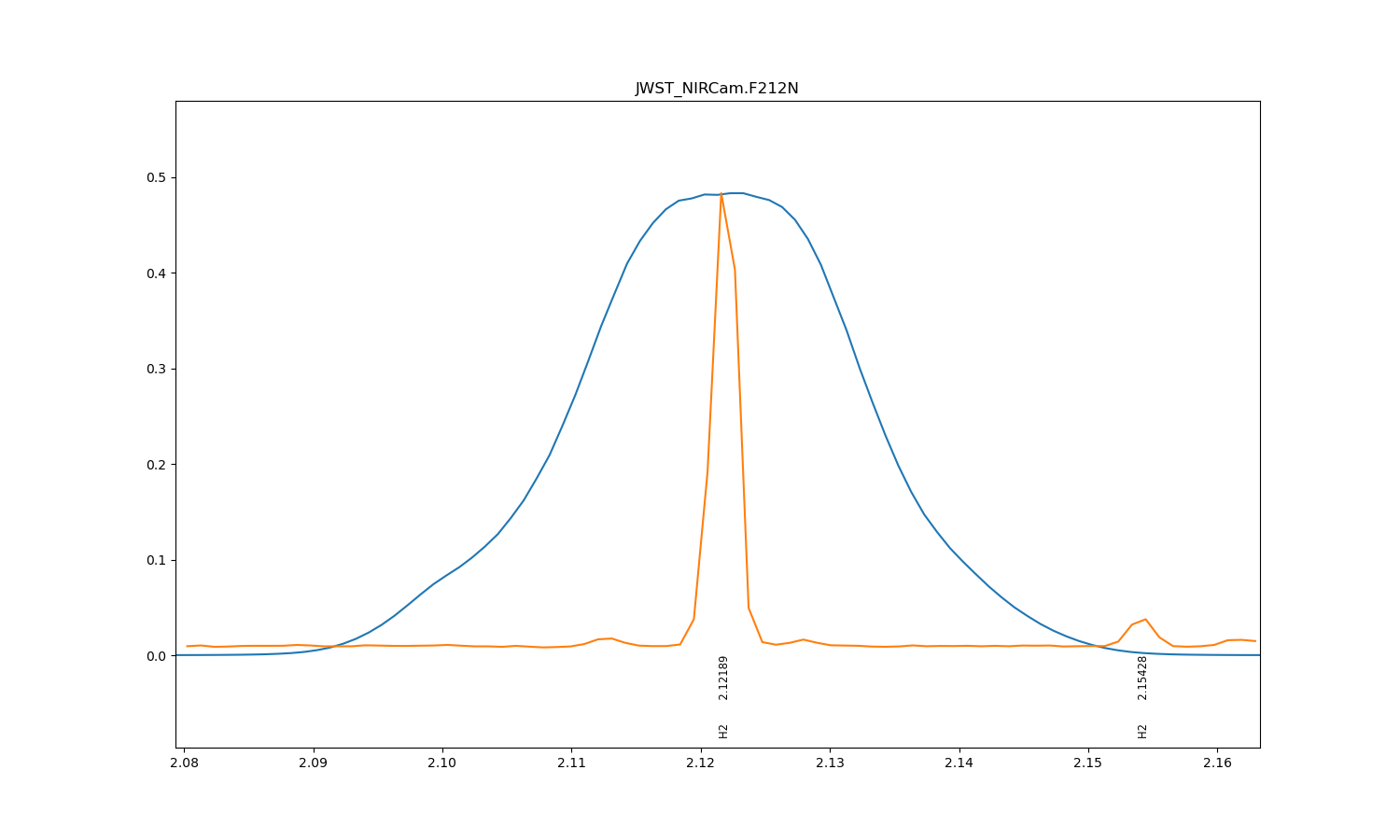}
    \includegraphics[trim={3.0cm 2.cm 3.3cm 2.2cm},clip, width=0.50\textwidth]{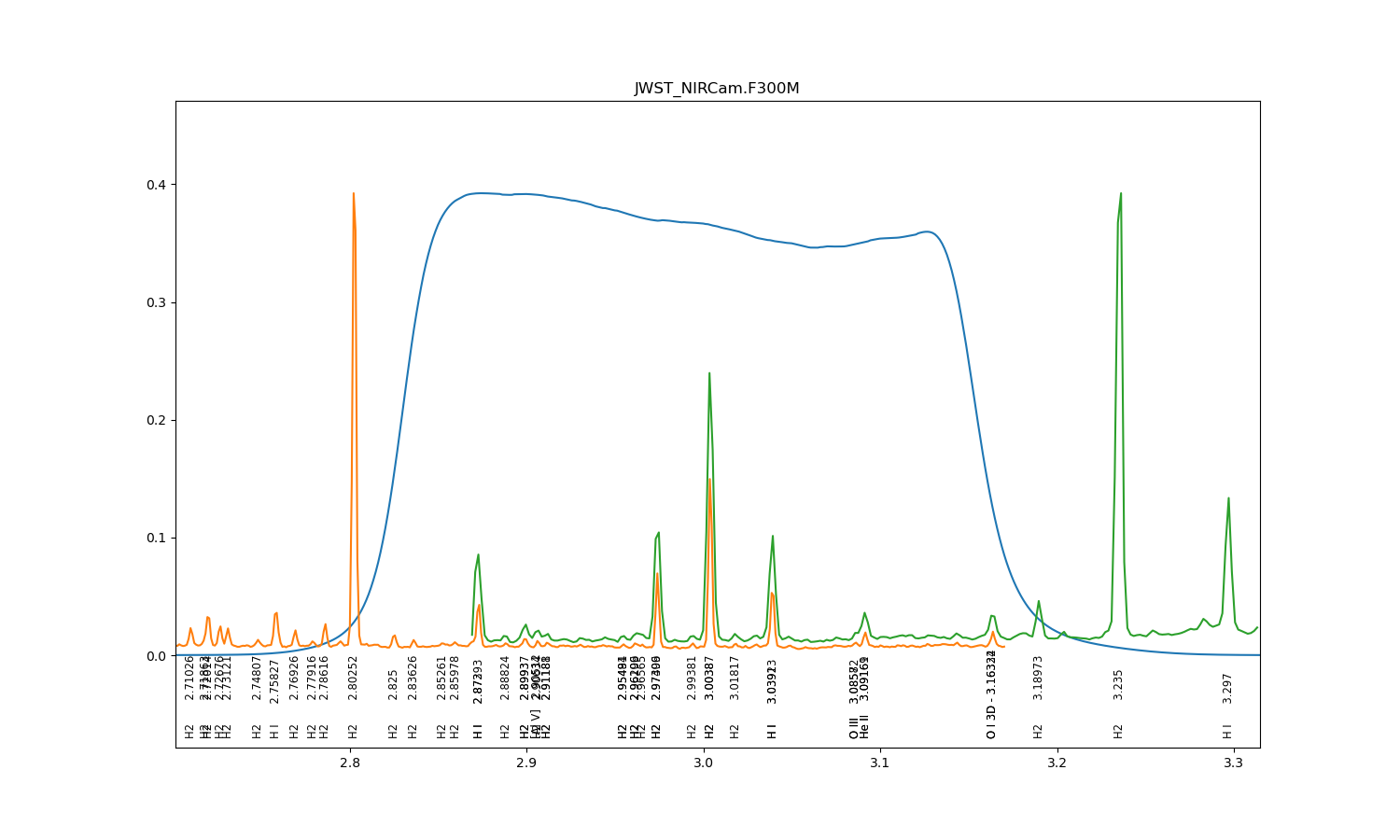}
    \includegraphics[trim={3.0cm 2.cm 3.3cm 2.2cm},clip, width=0.50\textwidth]{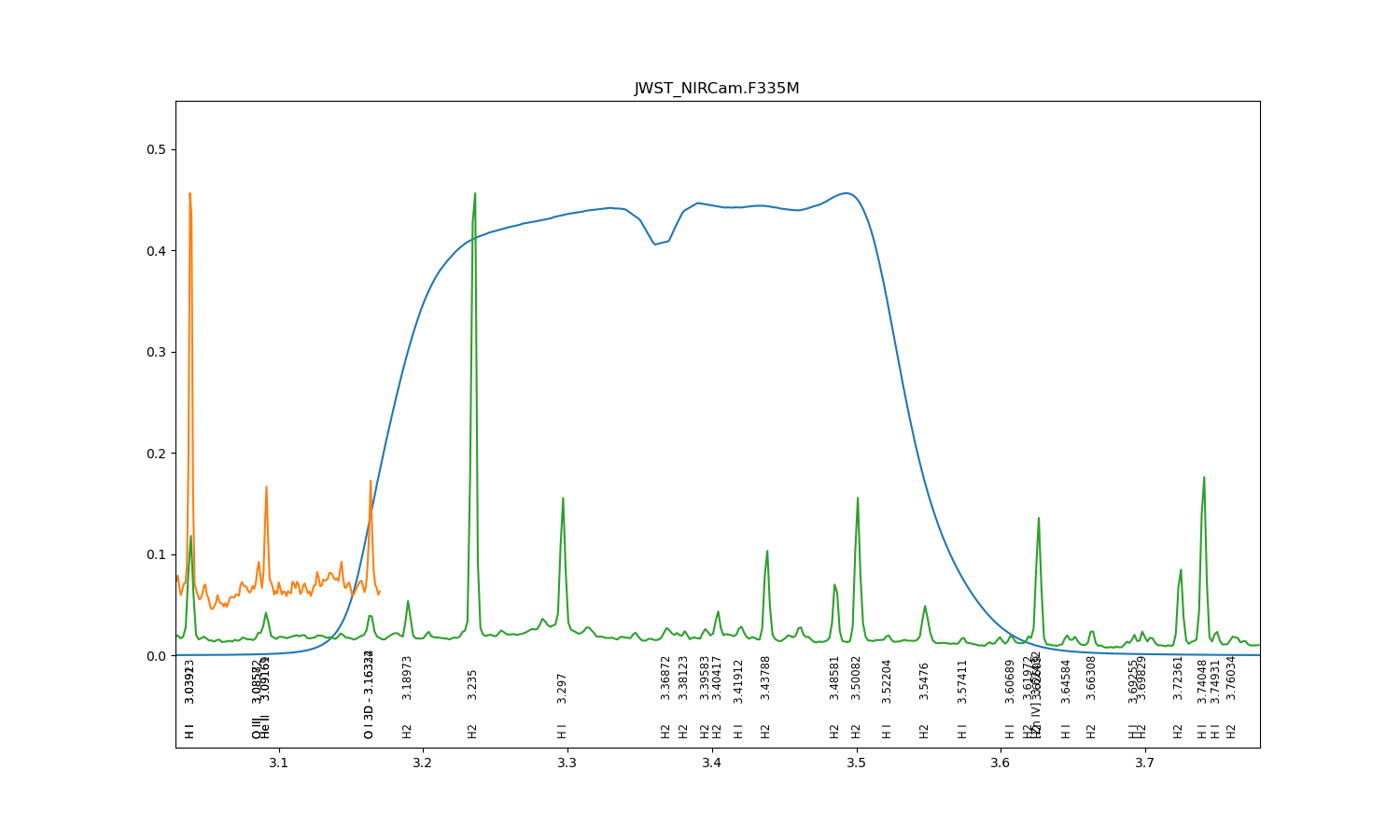}
       \end{minipage}\hfill
\caption{ The NIRCam filter transmission curves with NIRSpec spectra from the Northern part of the Ring. Line identifications from van Hoof et al. (in prep.) are included.
}
  \label{filter-NIRCam} 
\end{figure*}

Figures\,\ref{filter-NIRCam} -- \ref{filter-MIRI-2}  show the filter transmission curves 
\footnote{\url{https://jwst-docs.stsci.edu/jwst-near-infrared-camera/nircam-instrumentation/nircam-filters}}
\footnote{\url{https://jwst-docs.stsci.edu/jwst-mid-infrared-instrument/miri-instrumentation/miri-filters-and-dispersers}}
overlaid with NIRSpec and MIRI MRS spectra of the north region.

\begin{figure*}
  \begin{minipage}[c]{1\textwidth}
    \includegraphics[trim={3.0cm 2.cm 3.3cm 2.2cm},clip, width=0.50\textwidth]{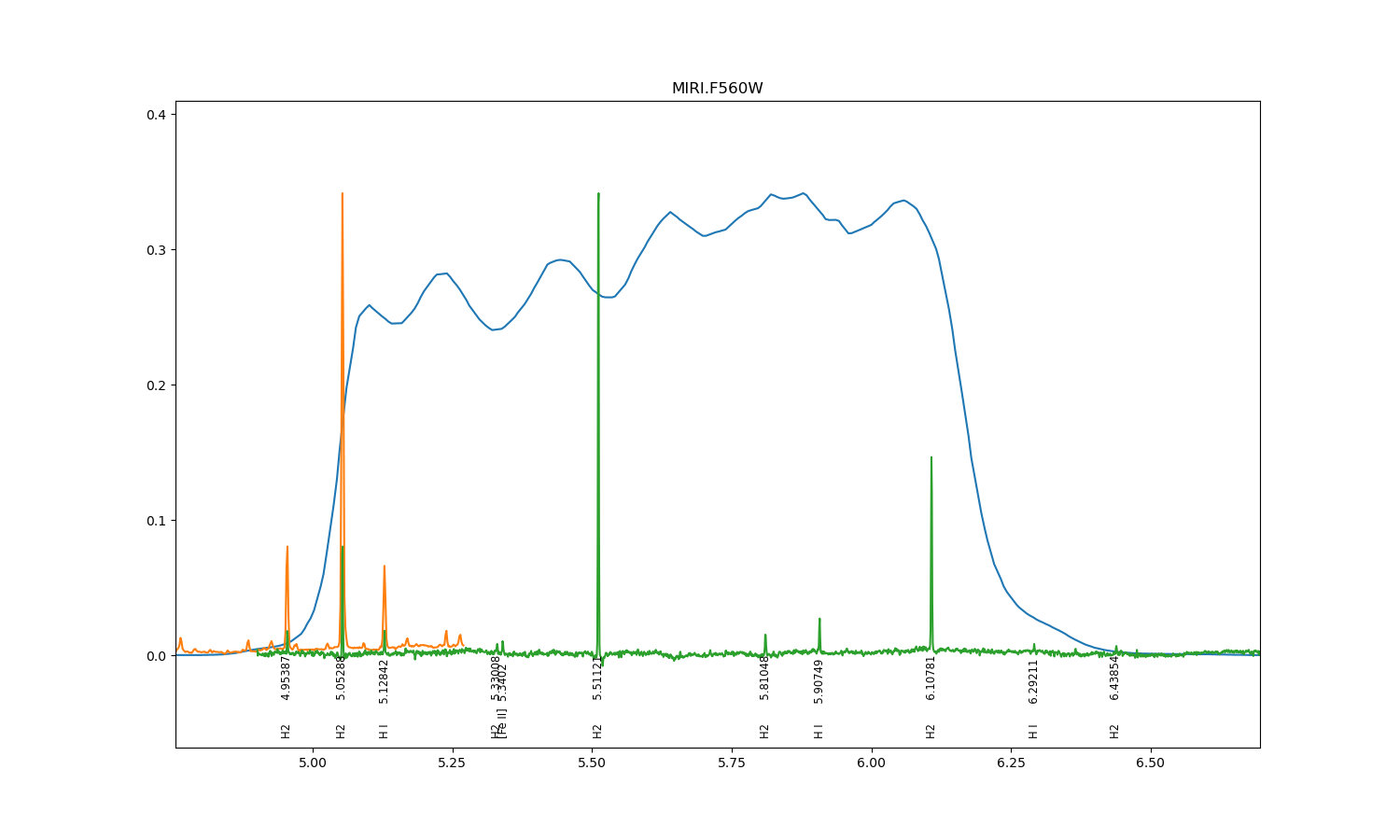}
    \includegraphics[trim={3.0cm 2.cm 3.3cm 2.2cm},clip, width=0.50\textwidth]{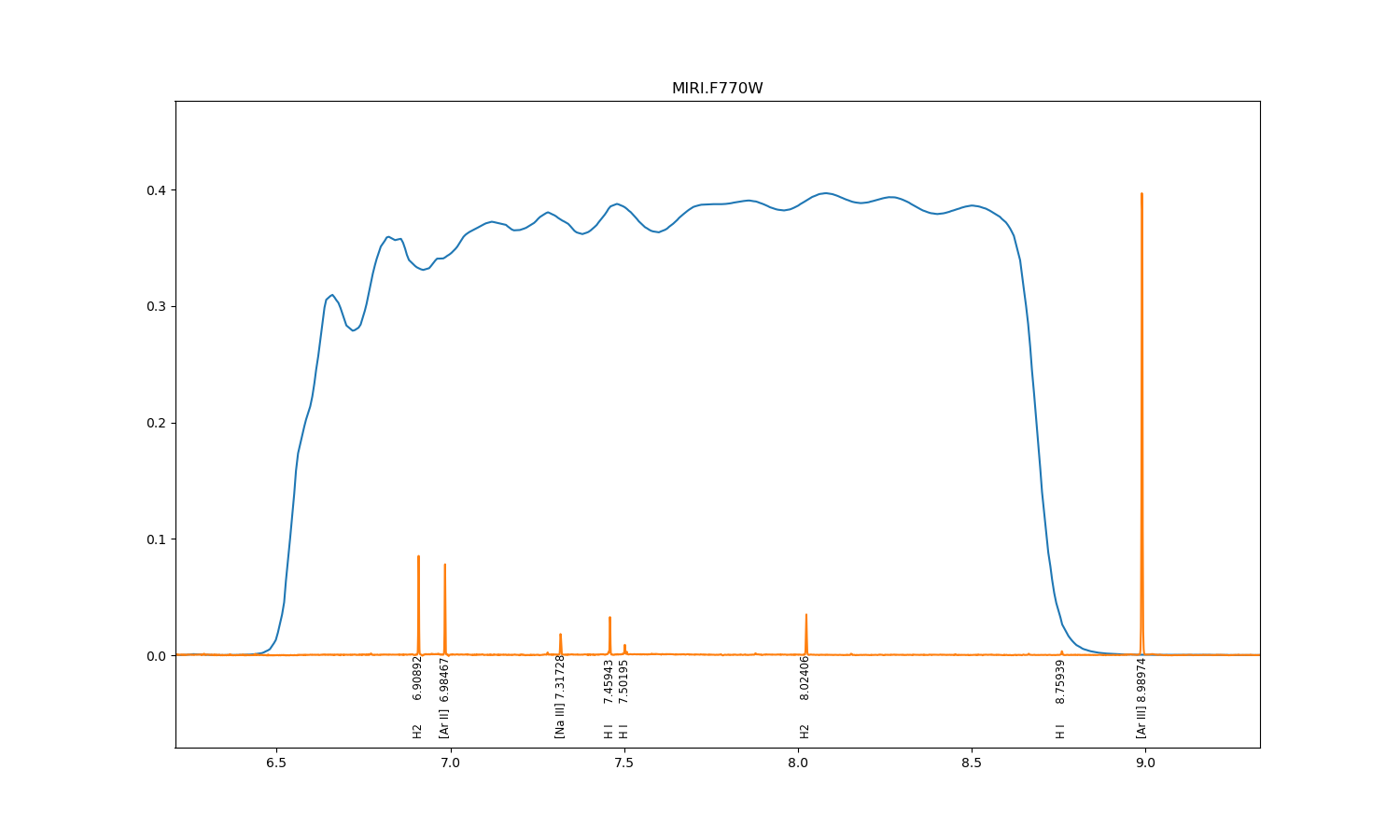}
    \includegraphics[trim={3.0cm 2.cm 3.3cm 2.2cm},clip, width=0.50\textwidth]{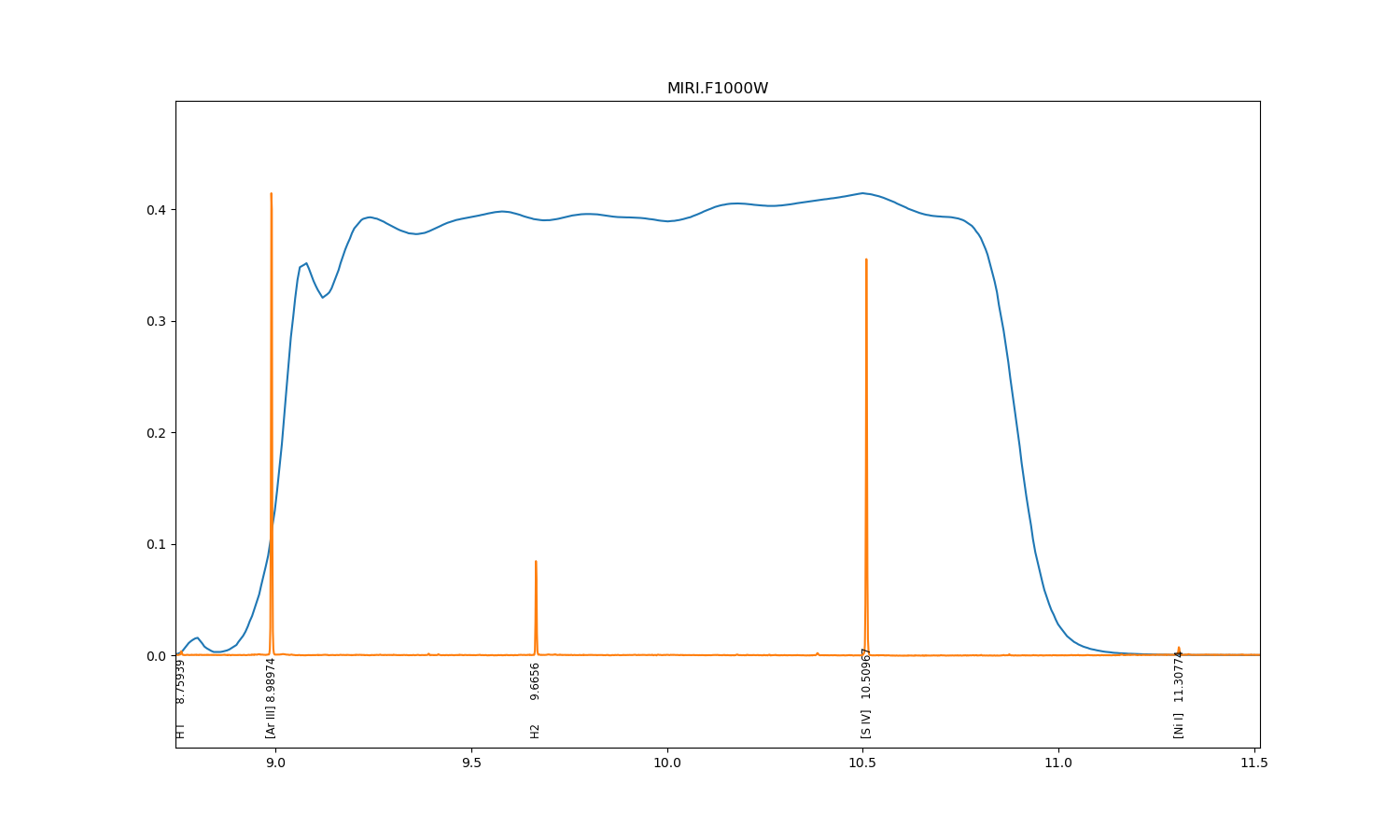}
    \includegraphics[trim={3.0cm 2.cm 3.3cm 2.2cm},clip, width=0.50\textwidth]{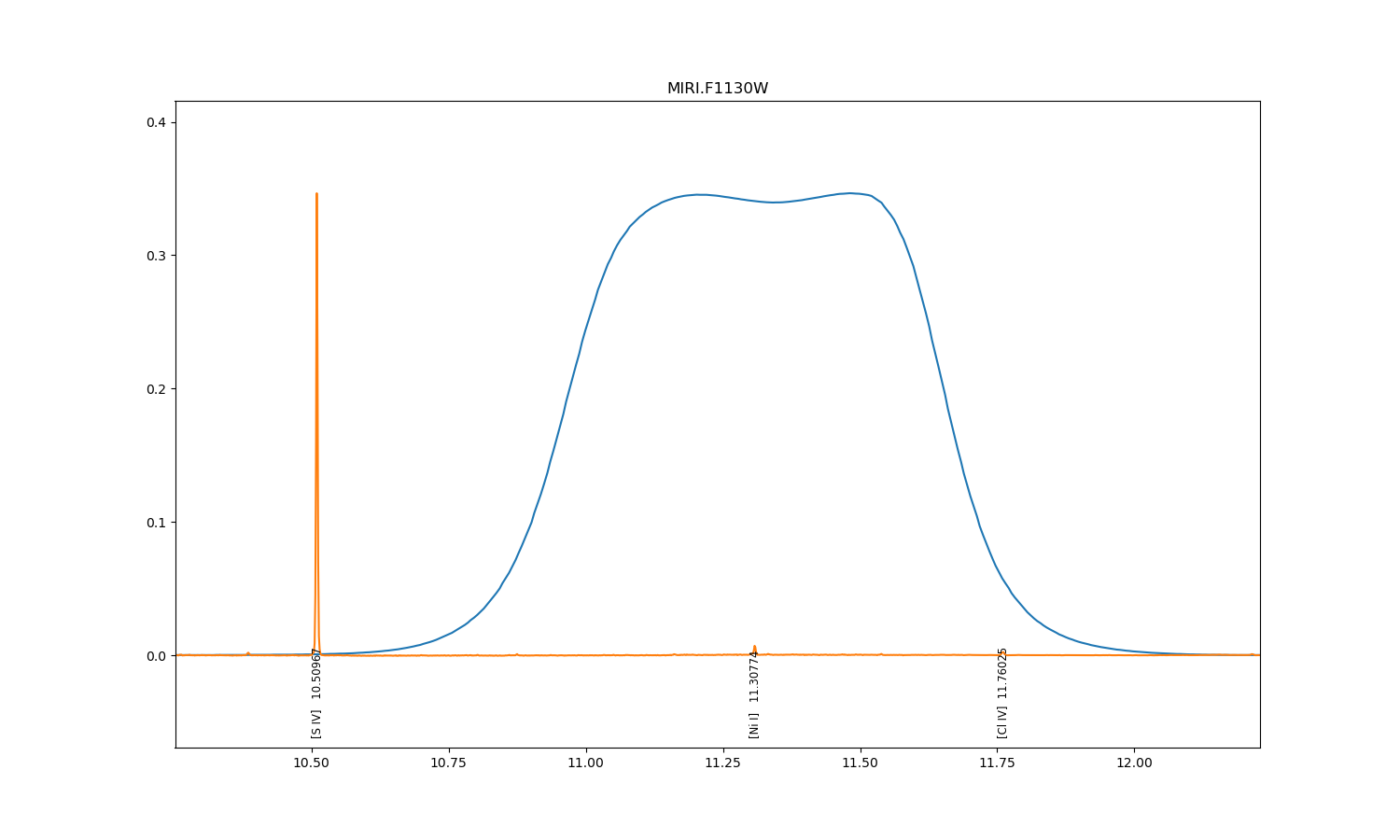}
       \end{minipage}\hfill
\caption{ The same as Fig.~\ref{filter-NIRCam}, but MIRI filters with MIRI MRS spectra 
}
  \label{filter-MIRI-1} 
\end{figure*}
\begin{figure*}
\ContinuedFloat
  \begin{minipage}[c]{1\textwidth}
    \includegraphics[trim={3.0cm 2.cm 3.3cm 2.2cm},clip, width=0.50\textwidth]{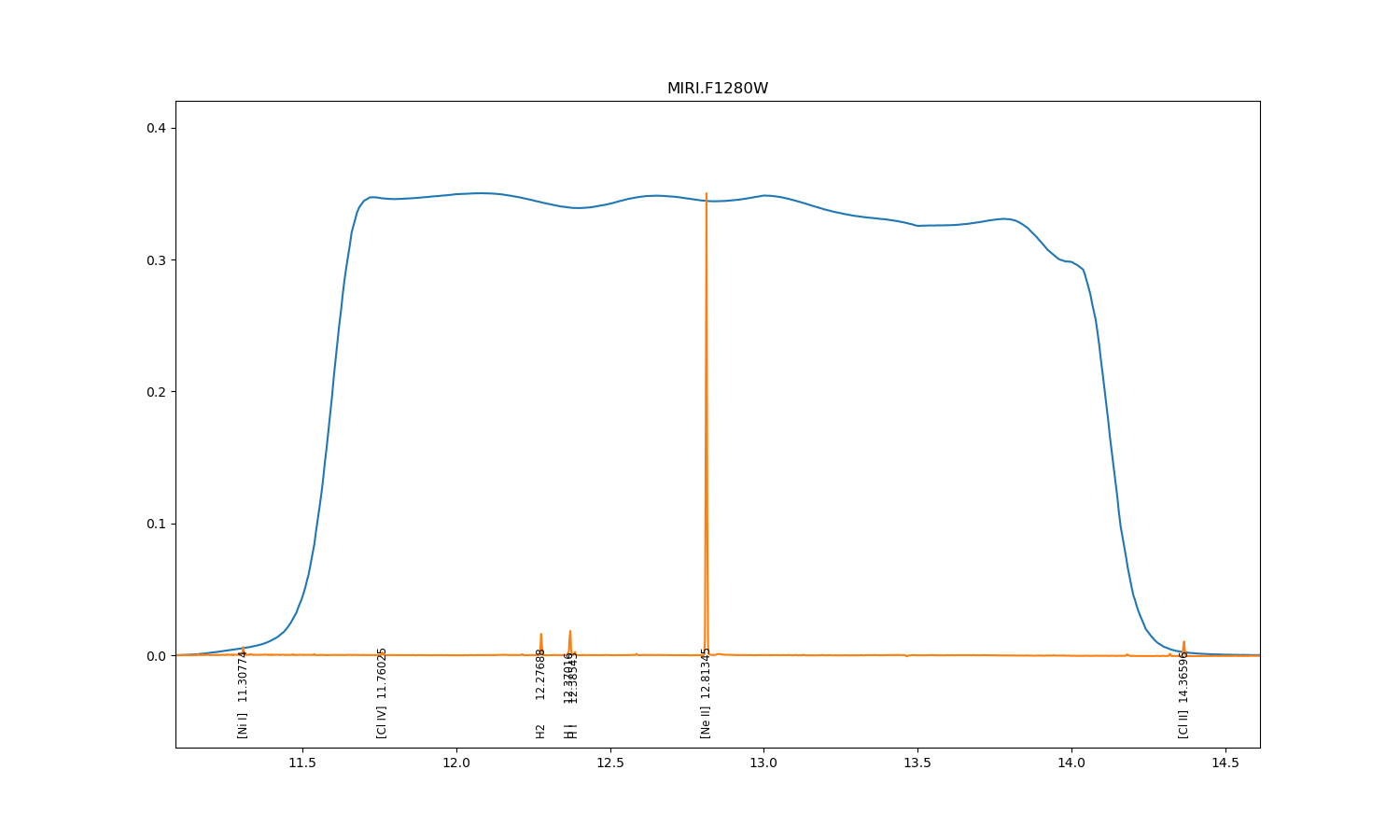}
    \includegraphics[trim={3.0cm 2.cm 3.3cm 2.2cm},clip, width=0.50\textwidth]{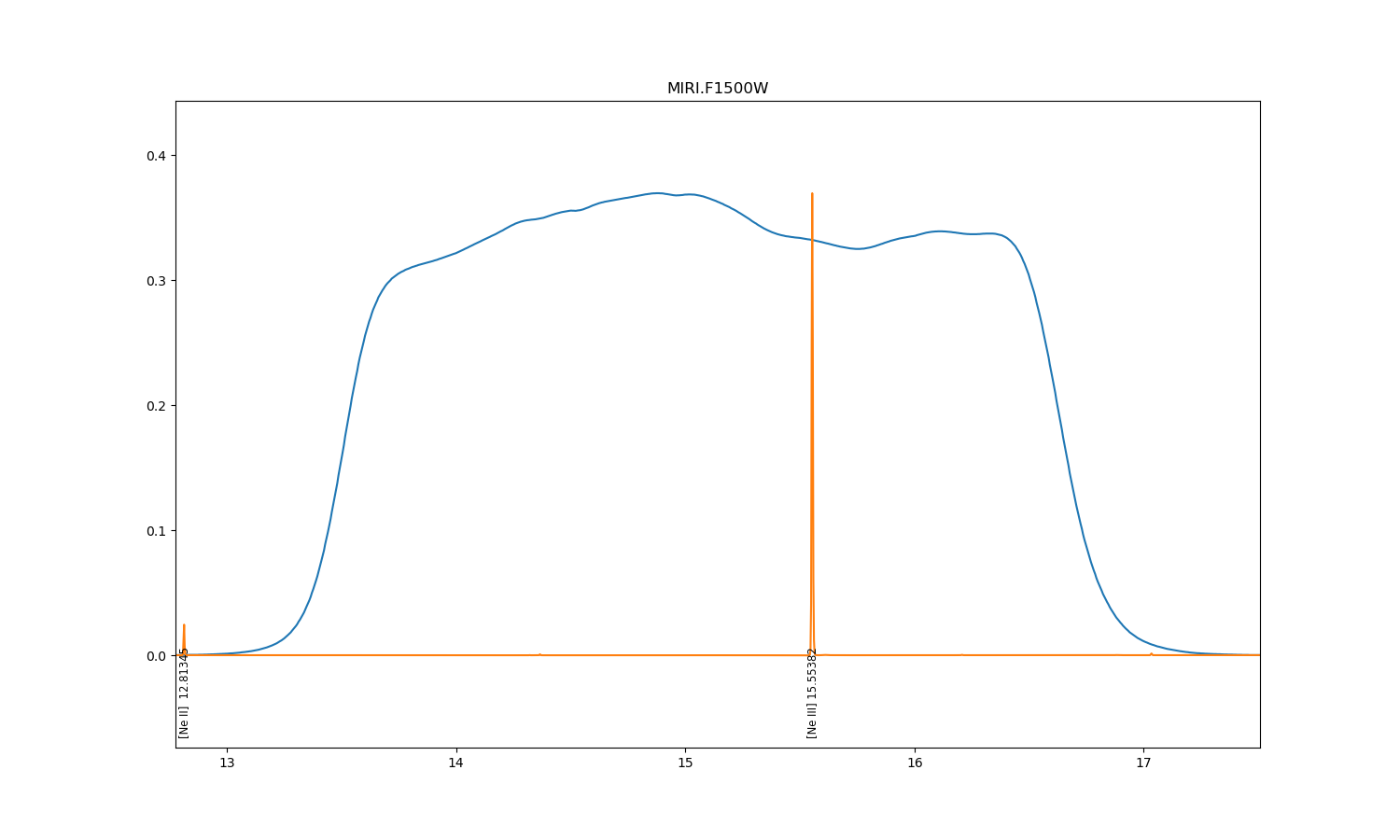}
    \includegraphics[trim={3.0cm 2.cm 3.3cm 2.2cm},clip, width=0.50\textwidth]{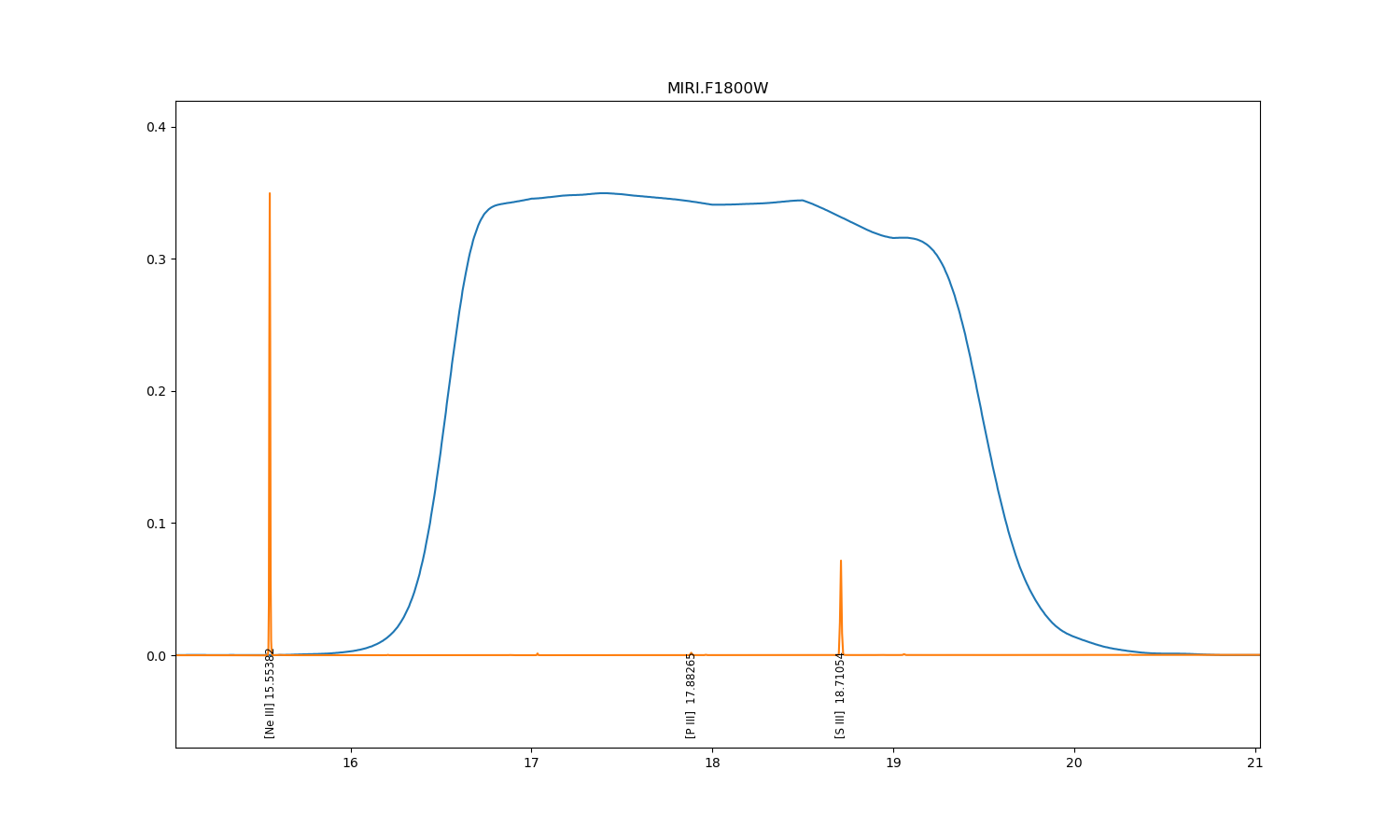}
    \includegraphics[trim={3.0cm 2.cm 3.3cm 2.2cm},clip, width=0.50\textwidth]{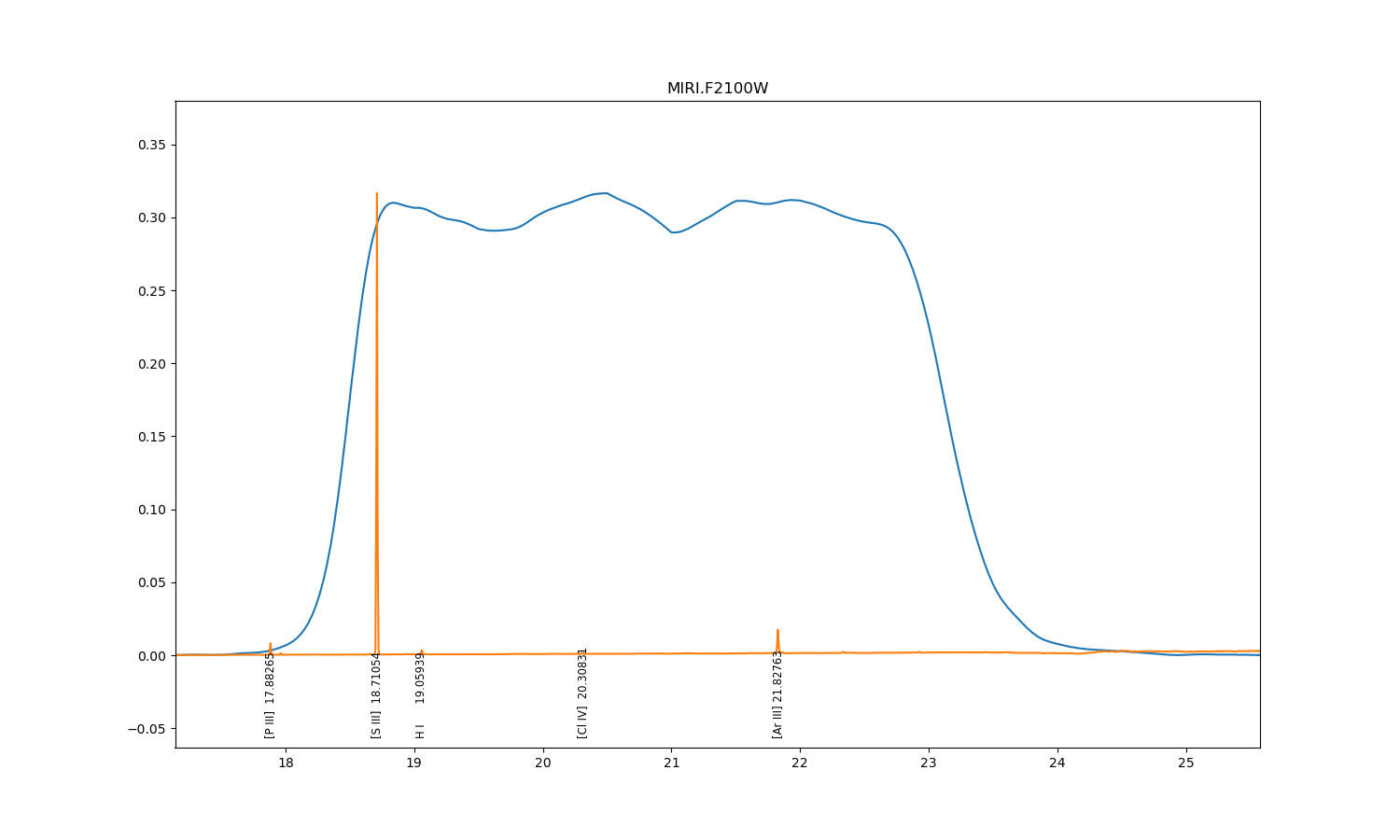}
    \includegraphics[trim={3.0cm 2.cm 3.3cm 2.2cm},clip, width=0.50\textwidth]{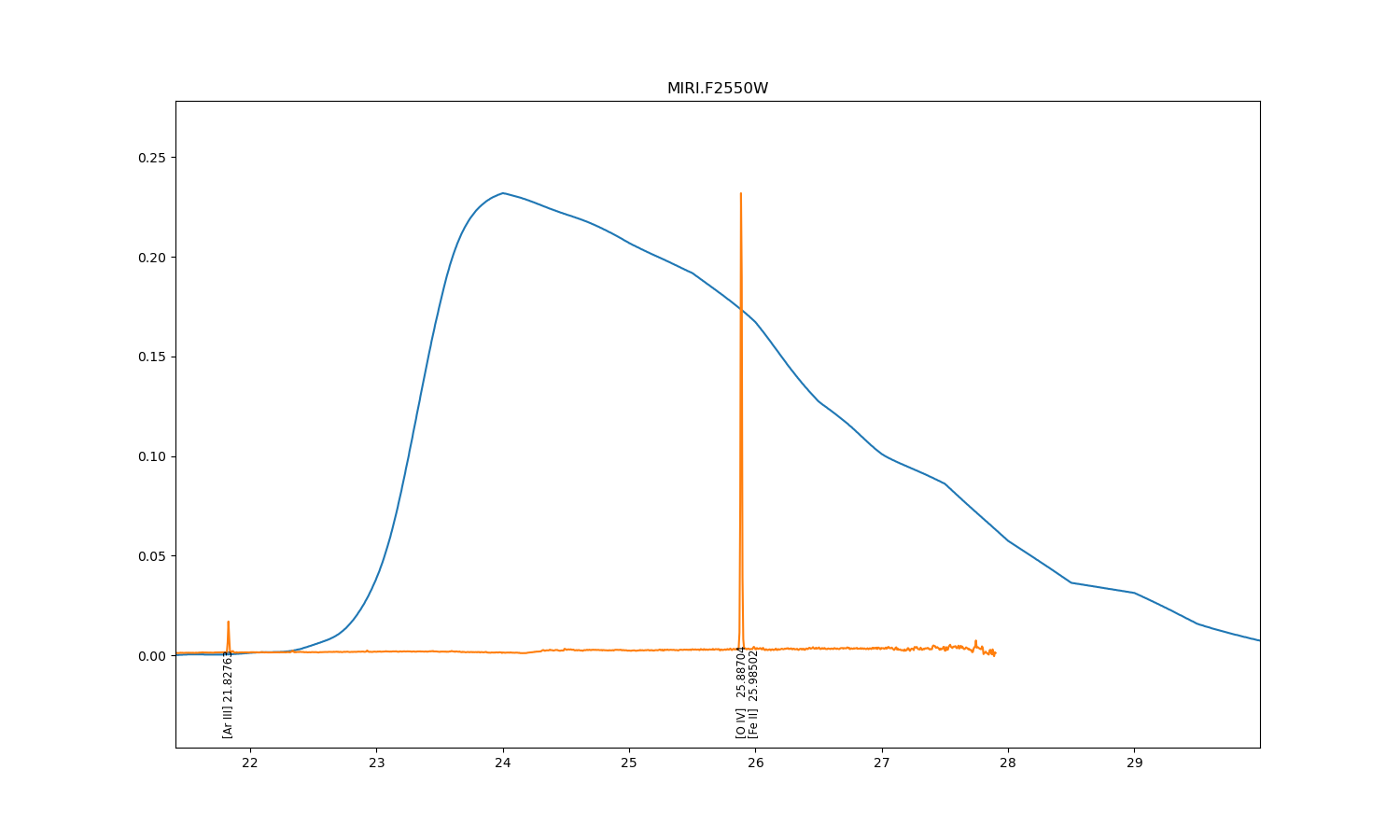}
       \end{minipage}\hfill
\caption{ (continued)
}
  \label{filter-MIRI-2} 
\end{figure*}

\begin{figure}
  \begin{minipage}[c]{1\textwidth}
    \includegraphics[trim={0.2cm 0.2cm 0.2cm 0.2cm},clip, width=0.5\textwidth]{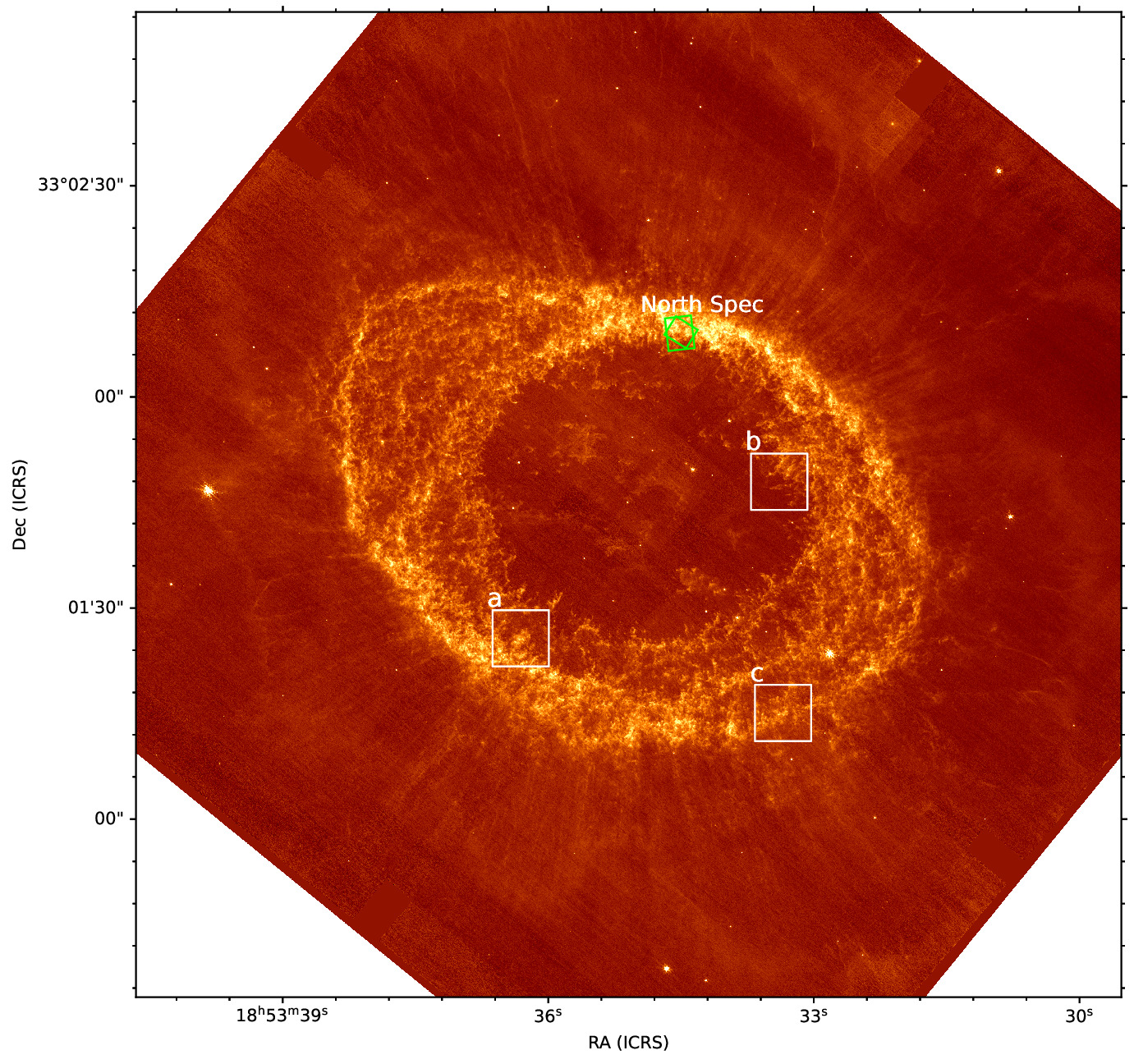}
      \end{minipage}\hfill
\caption{ Locations of globules plotted in Figure~\ref{globules} indicated by white boxes. The green boxes show the region where the northern spectra were taken. These spectra were used for filter transmission curves (Figs.~\ref{filter-NIRCam} and \ref{filter-MIRI-1} and \ref{filter-MIRI-2}) and contributions of lines into the filters Table~\ref{observing_log}. The smaller green box is for NIRSpec and the larger green box is for MIRI/MRS, both representing the smallest FoVs (fields of view) of these two instruments, as their FoVs have wavelength dependence.
}
  \label{globules-locations} 
\end{figure}

\begin{figure}
  \begin{minipage}[c]{1\textwidth}
    \includegraphics[trim={3.5cm 1.5cm 3cm 1cm},clip, width=0.5\textwidth]{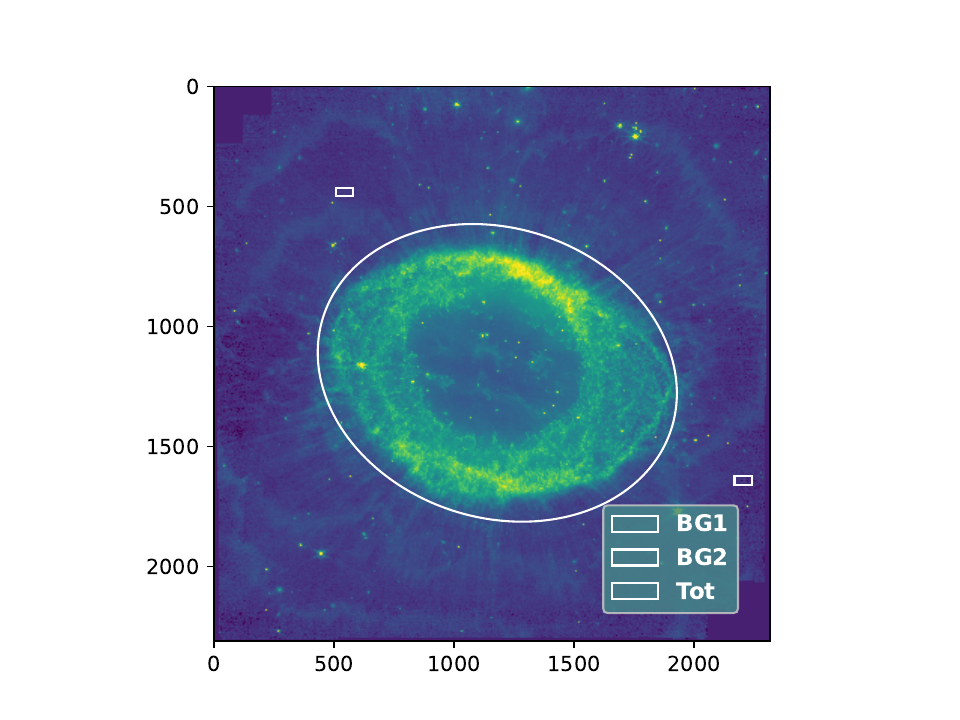}
      \end{minipage}\hfill
\caption{ 
Image showing the area used for the flux density (Tot) and the background regions (BG1 and BG2). 
The orientation is the same as Fig.~\ref{NIRCam-threecolor}
  \label{SED-locations} }
\end{figure}

\bsp	
\label{lastpage}
\end{document}